\documentclass[times,authoryear]{elsarticle}

\usepackage{jasr}
\usepackage{framed,multirow}

\usepackage{amssymb}
\usepackage{latexsym}

\usepackage[switch]{lineno}

\usepackage{url}
\usepackage{xcolor}
\definecolor{newcolor}{rgb}{.8,.349,.1}

\usepackage[citebordercolor=white]{hyperref}

\usepackage{acro}
\DeclareAcronym{TLE}{
    short = TLE,
    long = Two-Line Element Set
}

\DeclareAcronym{S-TLE}{
    short = S-TLE,
    long = Synthetic Two-Line Element Set
}

\DeclareAcronym{JPL}{
    short = JPL,
    long = Jet Propulsion Laboratory
}

\DeclareAcronym{ESA}{
    short = ESA,
    long = European Space Agency
}

\DeclareAcronym{ESA_Imperial}{
    short = ESA,
    long = Early Stage Assessment
}

\DeclareAcronym{LSR}{
    short = LSR,
    long = Late Stage Review
}

\DeclareAcronym{NASA}{
    short = NASA,
    long = National Aeronautics and Space Administration
}

\DeclareAcronym{IoS}{
    short = IoS,
    long = Interruption of Studies
}

\DeclareAcronym{PA}{
    short = PA,
    long = Polynomial Algebra
}

\DeclareAcronym{PAMC}{
    short = PAMC,
    long = Polynomial Algebra Monte Carlo
}

\DeclareAcronym{UQ}{
    short = UQ,
    long = Uncertainty Quantification
}

\DeclareAcronym{LEO}{
    short = LEO,
    long = Low Earth Orbit
}

\DeclareAcronym{MEO}{
    short = MEO,
    long = Medium Earth Orbit
}

\DeclareAcronym{GEO}{
    short = GEO,
    long = Geostationary Earth Orbit
}

\DeclareAcronym{GSO}{
    short = GSO,
    long = Geosynchronous Earth Orbit
}

\DeclareAcronym{EqOE}{
    short = EqOE,
    long = Equinoctial Orbital Element
}

\DeclareAcronym{GEqOE}{
    short = GEqOE,
    long = Generalised Equinoctial Orbital Element
}

\DeclareAcronym{AEqOE}{
  short = AEqOE,
  long  = Alternate Equinoctial Orbital Element
}

\DeclareAcronym{J2EqOE}{
  short = J\textsubscript{2}EqOE,
  long  = J\textsubscript{2} Equinoctial Orbital Element
}

\DeclareAcronym{GPS}{
    short = GPS,
    long = Global Positioning System
}

\DeclareAcronym{GNSS}{
    short = GNSS,
    long = Global Navigation Satellite System
}

\DeclareAcronym{WAAS}{
    short = WAAS,
    long = Wide Area Augementation System
}

\DeclareAcronym{CDDIS}{
    short = CDDIS,
    long = Crustal Dynamics Data Information System
}

\DeclareAcronym{KF}{
    short = KF,
    long = Kalman Filter
}

\DeclareAcronym{EKF}{
    short = EKF,
    long = Extended Kalman Filter
}

\DeclareAcronym{UKF}{
    short = UKF,
    long = Unscented Kalman Filter
}

\DeclareAcronym{DA}{
    short = DA,
    long = Differential Algebra
}

\DeclareAcronym{CPU}{
    short = CPU,
    long = Central Processing Unit
}

\DeclareAcronym{GPU}{
    short = GPU,
    long = Graphics Processing Unit
}

\DeclareAcronym{SRP}{
    short = SRP,
    long = Solar Radiation Pressure
}

\DeclareAcronym{STM}{
    short = STM,
    long = State Transition Matrix,
    long-plural-form = State Transition Matrices
}

\DeclareAcronym{RTN}{
  short = RTN,
  long = {Radial, Transverse, Normal},
  tag = abbrev
}

\DeclareAcronym{GP}{
  short = GP,
  long = General Perturbation,
  tag = abbrev
}

\DeclareAcronym{SP}{
  short = SP,
  long = Special Perturbation,
  tag = abbrev
}

\DeclareAcronym{SSA}{
  short = SSA,
  long = Space Situational Awareness,
  tag = abbrev
}

\DeclareAcronym{SDA}{
  short = SDA,
  long = Space Domain Awareness,
  tag = abbrev
}

\DeclareAcronym{USSF}{
  short = USSF,
  long = United States Space Force,
  tag = abbrev
}

\DeclareAcronym{CDM}{
  short = CDM,
  long = Conjunction Data Message,
  tag = abbrev
}

\DeclareAcronym{NORAD}{
  short = NORAD,
  long = North American Aerospace Defense Command,
  tag = abbrev
}

\DeclareAcronym{ILRS}{
  short = ILRS,
  long = International Lasing Ranging Service,
  tag = abbrev
}

\DeclareAcronym{LAGEOS}{
  short = LAGEOS,
  long = Laser Geodynamics Satellite,
}

\DeclareAcronym{SLR}{
  short = SLR,
  long = Satellite Laser Ranging,
}

\DeclareAcronym{OD}{
  short = OD,
  long = Orbit Determination,
}

\DeclareAcronym{P-OD}{
  short = P-OD,
  long = Pseudo-Orbit Determination,
}

\DeclareAcronym{EoM}{
    short = EoM,
    long = Equation of Motion,
    long-plural-form = Equations of Motion
}

\DeclareAcronym{XML}{
    short = XML,
    long = Extensible Markup Language
}

\DeclareAcronym{KVN}{
    short = KVN,
    long = Key-value Notation
}

\DeclareAcronym{JSON}{
    short = JSON,
    long = JavaScript Object Notation
}

\DeclareAcronym{RSO}{
    short = RSO,
    long = Resident Space Object
}

\DeclareAcronym{UT}{
    short = UT,
    long = Unscented Transform
}

\DeclareAcronym{CA}{
    short = CA,
    long = Conjunction Assessment
}

\DeclareAcronym{RMSE}{
    short = RMSE,
    long = Root-Mean-Square Error
}

\DeclareAcronym{ML}{
    short = ML,
    long = Machine Learning
}

\DeclareAcronym{LSODAR}{
    short = LSODAR,
    long = Livermore Solver for Ordinary Differential Equations with Automatic Method Switching for Stiff and Non-stiff Problems and with Root-finding,
}

\DeclareAcronym{TEME}{
    short = TEME,
    long = True Equator Mean Equinox,
}

\DeclareAcronym{GCRF}{
    short = GCRF,
    long = Geocentric Celestial Reference Frame,
}

\DeclareAcronym{IVP}{
    short = IVP,
    long = Initial Value Problem,
}
\usepackage{amsmath}
\usepackage{booktabs}
\usepackage{csquotes}
\usepackage[capitalise]{cleveref}
\usepackage{siunitx}
\usepackage{tikz}
\usepackage{caption}
\usepackage{subcaption}
\usepackage{placeins}

\journal{Advances in Space Research}

\begin{document}

\verso{Max I. Hallgarten La Casta \textit{et al.}}

\begin{frontmatter}

\title{Debiasing of Two-Line Element Sets for Batch Least Squares Pseudo-Orbit Determination in MEO and GEO}

\date{December 19, 2024}

\author[1]{Max I. \snm{Hallgarten La Casta}\corref{cor1}}
\cortext[cor1]{Corresponding author. Email: m.hallgarten-la-casta21@imperial.ac.uk}
\author[1]{Davide \snm{Amato}\fnref{fn1}}
\fntext[fn1]{Email: d.amato@imperial.ac.uk}

\affiliation[1]{organization={Imperial College London},
                addressline={Exhibition Road},
                city={London},
                postcode={SW7 2AZ},
                country={United Kingdom}}


\begin{abstract}

The availability of accurate and timely state predictions for objects in near-Earth orbits is becoming increasingly important due to the growing congestion in key orbital regimes. The \ac{TLE} catalogue remains, to this day, one of the few publicly-available, comprehensive sources of near-Earth object ephemerides. At the same time, \acp{TLE} are affected by measurement noise and are limited by the low accuracy of the SGP4 theory, introducing significant uncertainty into state predictions. Previous literature has shown that filtering \acp{TLE} with batch least squares methods can yield significant improvements in long-term state prediction accuracy. However, this process can be highly sensitive to \ac{TLE} quality which can vary throughout the year. In this study, it is shown that either extended-duration fit windows of the order of months, or the removal of systematic biases in along-track position prior to state estimation can produce significant reductions in post-fit position errors. Simple models for estimating these systematic biases are shown to be effective without introducing the need for high-complexity \ac{ML} models. Furthermore, by establishing a \acs{TLE}-based error metric, the need for high accuracy ephemerides is removed when creating these models. For selected satellites in the \ac{MEO} regime, post-fit position errors are reduced by up to \SI{80}{\%}, from approximately \SIrange{5}{1}{\kilo\metre}; meanwhile, for selected satellites in the \ac{GEO}/\ac{GSO} regime, large oscillations in post-fit position error can be suppressed.

\end{abstract}


\end{frontmatter}


\acresetall

\section{Introduction}\label{s:introduction}

The rise of the \enquote{New Space} era, driven by the growth in the number of commercial actors in the space domain, has led to a significant increase in the near-Earth \ac{RSO} population over the past decade, including year-on-year growth in the number of launches and payloads~\citep{noauthor_esas_2023}.

The growth in the \ac{RSO} population has been seen primarily in the \ac{LEO} and \ac{GEO} regimes, therefore the majority of these objects are concentrated into the same limited regions. The risk of collisions between \acp{RSO} continues to grow, highlighting the importance of improving the capacity and capabilities of \ac{CA} systems. These are highly dependent on accurate and precise state estimates to remain effective. With future exploration and development, the cislunar environment will become increasingly relevant, underscoring the need to understand various different orbital regimes.

Many operators have high precision state estimates for their own spacecraft based on their own on-board sensors, however they remain dependent on governmental and commercial \ac{SSA} organisations, and data sharing with other operators, for state estimates of other objects. The financial cost of these services mean that smaller operators and academia remain reliant on publicly available sources, primarily \acp{TLE} published by the \ac{USSF}.

One of the greatest advantages of \acp{TLE} is that they are freely available for a significant proportion of \acp{RSO} currently (and formerly) in near-Earth orbit. They represent an unparalleled source of data for \ac{RSO} state vectors. Nevertheless, they are subject to a number of limitations which are difficult to overcome, primarily their limited accuracy when predicting states. This is due to the analytical models used by the format (SGP4 and SDP4, now commonly referred to jointly as SGP4) which use simplified perturbation models with limited fidelity~\citep{hoots_history_2004}. This is despite the fact that \acp{TLE} themselves are generated by fitting the models to higher-order predictions made by direct \ac{OD}~\citep{hejduk_catalogue-wide_2013}. Additionally, the lack of uncertainty information for \acp{TLE} mean that their accuracy, even at epoch, remains effectively unknown. The result is a widely used data format with high availability across the orbital catalogue but with unknown or poor accuracy, making them unsuitable for precision applications.

The limited fidelity of the SGP4 model was identified as an issue as \acp{TLE} became more widely used, particularly when considering \enquote{the requirements and problems in modern space mission operations}~\citep{conkey_assessing_2022}. For example, the stringent requirements on accuracy for \ac{CA} remain unsatisfied by traditional \acp{TLE} which are only sufficient for \enquote{providing a first-order screening for collision alerting}~\citep{chao_applied_2018}. The limitations of traditional \acp{TLE} and SGP4 led to the development of the SGP4-XP model, a new algorithm for propagating \acp{TLE} with greater precision, albeit with a reduction in computational performance~\citep{conkey_assessing_2022}. An extension of SGP4, this updated model includes improved geopotential and resonance modelling, higher order lunisolar perturbations, a more detailed atmospheric model, and improved \ac{SRP} modelling~\citep{conkey_assessing_2022,payne_improvements_2022,holincheck_improved_2021}. It should be noted that \acp{TLE} using SGP4-XP, designated as Type~4 ephemerides, are not compatible with the original SGP4 model, nor vice versa~\citep{conkey_assessing_2022}. This is a product of the models being based on different averaging theories and, hence, they correspond to different dynamical systems.

Binaries for SGP4-XP are available from \ac{USSF}; however, mean elements compatible with the SGP4-XP theory have not been publicly distributed as of the date of writing. Nevertheless, testing with synthetically generated Type~4 \acp{TLE} has demonstrated significant improvements in prediction accuracy for objects in the \ac{MEO} and \ac{GEO}/\ac{GSO} regimes~\citep{conkey_assessing_2022}. One feature, particularly critical for \ac{P-OD}, is the \enquote{more realistic dynamics coming from the SGP4-XP algorithm [yielding] a structure in the error which is more Gaussian than that of the classical SGP4}~\citep{conkey_assessing_2022}. This includes a significant reduction in along-track error~\citep{payne_improvements_2022} which is the primary component of classical \ac{TLE} error. Since Type~4 \acp{TLE} remain unavailable publicly, methods for reducing \ac{TLE} biases remain an important area of investigation for deriving higher precision ephemerides.

Previous literature has demonstrated that it is possible to use \acp{TLE} in conjunction with higher fidelity orbital propagators to produce more accurate state estimations though \ac{P-OD}. This is the process of conducting \ac{OD} on products which themselves are the result of an \ac{OD} process. \cite{levit_improved_2011} conducted batch least squares fitting of a high-fidelity numerical model to \ac{TLE} pseudo-observations, improving the error rate for position prediction in \ac{LEO} from approximately \SIrange{100}{3000}{\metre/day} to \SIrange{50}{200}{\metre/day}. \cite{vallado_improved_2013} extended existing analyses by investigating the influence of different estimation parameters, concluding that uncertainty was largest in the along-track direction and that force model fidelity had a limited impact on results. \cite{bennett_improving_2012} demonstrated that relatively simple bias models can be used to improve \acs{TLE}-based \ac{P-OD} when fused with \ac{SLR} data. \cite{chen_research_2023}, in a study of \ac{P-OD} techniques on satellites in \ac{LEO}, showed that fit prediction accuracy can be improved by restricting sampling to within two orbital periods of each epoch. \cite{ly_correcting_2020} investigated using statistical and \ac{ML} techniques to find corrections for \ac{GPS} satellite \acp{TLE} to improve prediction, leading to \enquote{improved TLEs accuracy from 5~km to 1.5~km with a 90\% confidence level}. This study focused on correcting the along-track error present in \acp{TLE} with kernel regression techniques which revealed correlation between the error and both the Sun-Earth and Moon-Earth distances.

The objective of this study is to revisit past analyses of \ac{P-OD} on \acp{TLE} to confirm whether these methods are still effective in light of changes in \ac{TLE} behaviour over the past decade and, additionally, to investigate the ideal parameters when using batch least squares for achieving the best possible accuracy. We consider passive satellites, both with and without accurate reference ephemerides, to establish an error metric that is purely \acs{TLE}-based to enable analysis of any object in the \ac{TLE} catalogue.

The methodology for conducting \ac{P-OD} will be discussed in \cref{s:methodology}, followed by a presentation of the results in \cref{s:results}, including a discussion of considerations for the further development of bias models. \cref{s:conclusions} will provide a summary of the outcomes of the study and potential avenues for future development.

\section{Overview of Pseudo-Orbit Determination}\label{s:methodology}

\begin{figure}[htb]
    \centering

    \vspace{0.5\baselineskip}

    \begin{tikzpicture}

        \draw[blue, thick] (0,0.2) -- +(0, -0.4) node[below, align=center]{Fit \\ Start};
        \draw[blue, thick] (2.5,0.2) -- +(0, -0.4) node[below, align=center]{Fit \\ Midpoint};
        \draw[blue, thick] (5,0.0) -- +(0, 0.2);
        \draw[red, thick] (5,0.0) -- +(0, -0.2) node[below, align=center]{\textcolor{blue}{Fit End} \\ \textcolor{red}{Prediction Start}};
        \draw[red, thick] (15,0.2) -- +(0, -0.4) node[below, align=center]{Prediction \\ End};

        \draw[blue, thick] (0,0) -- (5,0);
        \draw[red, thick] (5,0) -- (15,0);

        \draw[black,|-|] (0, -1.5) -- node[below]{Fit Duration} (5, -1.5);
        \draw[black,|-|] (5, -1.5) -- node[below]{Prediction Duration} (15, -1.5);

    \end{tikzpicture}

    \vspace{0.5\baselineskip}

    \caption{Fit and prediction windows.}
    \label{f:windows}
\end{figure}
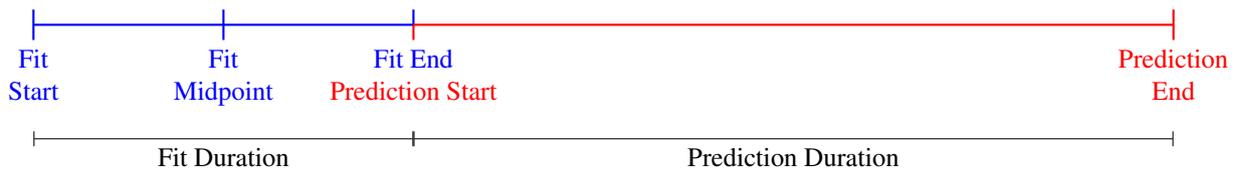

The \ac{P-OD} process can be split into two main components: fit and prediction. In the first, pseudo-observations generated from a set of \acp{TLE} are used to estimate the state of the satellite. This is followed by additional propagation in the prediction phase to evaluate the quality of the fit by comparing it against a reference \enquote{ground truth} set of states provided by either \ac{SLR} or \acp{TLE}. The fit and prediction windows are illustrated in \cref{f:windows} with key quantities including the fit start and end dates, followed immediately by the prediction window.

\subsection{Batch Least Squares}
The batch least squares estimator considers the problem of selecting model parameters which minimise the errors between states predicted from the model and a set of observed states. In this case, the model is one which represents the future state of a satellite as a function of initial state, physical parameters of the satellite, and prediction time. The following derivation summarises the method as described by \cite{vallado_fundamentals_2013} and \cite{tapley_statistical_2004}.

\subsubsection{Predictor}

The predictor is an \ac{IVP} where the state of an object can be predicted based on an initial state: 
\begin{equation}
    \dot{\mathbf{x}} = \mathbf{f}(\mathbf{x}, t) \text{,}
\end{equation}
\begin{equation}
    \mathbf{x}(t_0) = \mathbf{x}_0 \text{,}
\end{equation}
where $\mathbf{x}$ and $\dot{\mathbf{x}}$ are the state and state derivative, $t$ is time, $\mathbf{f}(\mathbf{x}, t)$ is the dynamical model of the predictor, and $t_0$ and $\mathbf{x}_0$ are the initial time and state respectively.

The state at a future time can be predicted by integrating with respect to time:
\begin{equation}
    \mathbf{y}^*(t) = \mathbf{x}_0 + \int_{t_0}^t \dot{\mathbf{x}} \ dt \text{,}
\end{equation}
where $\mathbf{y}^*(t)$ is the state at a future time $t$. Alternatively, this can be expressed as the solution of a linear system:
\begin{equation}\label{e:problem_yAx}
    \mathbf{y}^*(t) = \mathbf{A}(t, t_0) \mathbf{x}_0 \text{,}
\end{equation}
where $\mathbf{A}(t, t_0)$ is the partial derivatives matrix which maps between initial and predicted states. The derivatives matrix can be further defined as:
\begin{equation}
    \mathbf{A}(t, t_0) = \mathbf{H}(t) \mathbf{\Phi}(t,t_0) \text{,}
\end{equation}
where $\mathbf{H}(t)$ is the observation matrix at time $t$ which maps from state space to observation space, and $\mathbf{\Phi}(t,t_0)$ is the \ac{STM} which maps from the initial state at time $t_0$ to the predicted state at time $t$. In this case, the state and observation spaces were the same, therefore the observation matrix was simply identity.

By particularising \cref{e:problem_yAx} for $n$ epochs:
\begin{equation}
    \begin{aligned}
        \mathbf{y}^*(t_0) &= \mathbf{A}(t_0, t_0) \mathbf{x}_0 \text{,} \\
        \mathbf{y}^*(t_1) &= \mathbf{A}(t_1, t_0) \mathbf{x}_0 \text{,} \\
        \mathbf{y}^*(t_2) &= \mathbf{A}(t_2, t_0) \mathbf{x}_0 \text{,} \\
                          &\vdots                                       \\
        \mathbf{y}^*(t_n) &= \mathbf{A}(t_n, t_0) \mathbf{x}_0 \text{,}
    \end{aligned}    
\end{equation}
the individual linear systems can be concatenated into a single overall linear system:
\begin{equation}
    \begin{bmatrix}
        \left[ \mathbf{y}^*_0 \right] \\
        \left[ \mathbf{y}^*_1 \right] \\
        \left[ \mathbf{y}^*_2 \right] \\
        \vdots                        \\
        \left[ \mathbf{y}^*_n \right]
    \end{bmatrix}
    =
    \begin{bmatrix}
        \left[ \mathbf{A}_0 \right] \\
        \left[ \mathbf{A}_1 \right] \\
        \left[ \mathbf{A}_2 \right] \\
        \vdots                      \\
        \left[ \mathbf{A}_n \right]
    \end{bmatrix}
    \mathbf{x}_0
    \text{,}
\end{equation}
where:
\begin{equation}
    \mathbf{y}^*_i = \mathbf{y}^*(t_i) \text{,}
\end{equation}
\begin{equation}
    \mathbf{A}_i = \mathbf{A}(t_i, t_0) \text{.}
\end{equation}

The predicted state vectors, therefore, can be expressed as a function of an overall derivatives matrix and the initial state vector:
\begin{equation}
    \mathbf{y}^* = \mathbf{A} \mathbf{x}_0 \text{.}
\end{equation}

\subsubsection{Observer}
Object states are represented in \acp{TLE} by a number of parameters, including a set of mean elements, based on a development from theories by Brouwer and Kozai, among others~\citep{hoots_history_2004,chao_applied_2018}, and a ballistic coefficient. Propagation of these parameters to future times, and conversion to Cartesian state, is based on the SGP4 single-averaged theory. For orbits with a period greater than \SI{225}{minutes}, deep space perturbations are included to \enquote{account for lunar and solar gravitation as well as the resonance effects of Earth tesseral harmonics}~\citep{chao_applied_2018}.

Propagation with SGP4 can be expressed as an analytical expression:
\begin{equation}
    \mathbf{y}(t) = \mathbf{R} \: \mathbf{g}(\mathbf{z}_k, t) \text{,}
\end{equation}
\begin{equation}
    \mathbf{z}(t_k) = \mathbf{z}_k \text{,}
\end{equation}
where $\mathbf{y}(t)$ is the \ac{TLE} state at a future time $t$; $\mathbf{R}$ is the rotation matrix from \ac{TEME} to \ac{GCRF}; $\mathbf{g}(\mathbf{z}, t)$ represents SGP4; and $t_k$ and $\mathbf{z}_k$ are the epoch and state of the appropriate \ac{TLE} respectively, including both the mean elements and ballistic coefficient.

SGP4 conducts propagation in its set of mean elements which are converted then into Cartesian positions and velocities in \ac{TEME}. These are transformed into \ac{GCRF} for consistency with the predictor. Both the SGP4 propagation and frame transformations were provided by the Orekit spaceflight library~\citep{maisonobe_orekit_2010}.

Similar to the predictor, the observations are concatenated into a single column vector of states for $n$ epochs:
\begin{equation}
    \mathbf{y}
    =
    \begin{bmatrix}
        \mathbf{y}(t_0) \\
        \mathbf{y}(t_1) \\
        \mathbf{y}(t_2) \\
        \vdots          \\
        \mathbf{y}(t_n)
    \end{bmatrix}
    \text{.}
\end{equation}

Further details regarding pseudo-observation generation are provided in \cref{s:tle_set_sampling}.

\subsubsection{Least Squares solver}

The errors between predictions and observations are expressed as a set of residuals:
\begin{align}
    \mathbf{b} &= \mathbf{y} - \mathbf{y}^* \text{,}          \label{e:residual_noa} \\
               &= \mathbf{y} - \mathbf{A} \mathbf{x} \text{,} \label{e:residual}
\end{align}
where $\mathbf{b}$ is the residual vector; $\mathbf{y}$ is the vector of \mbox{(pseudo-)observations}; $\mathbf{y}^*$ is a vector of predictions; $\mathbf{A} = \partial \mathbf{y}^* / \partial \mathbf{x}$ is the partial derivatives matrix of the system, in this case computed with finite differences as described by \cite{vallado_fundamentals_2013}; and $\mathbf{x}$ redefined to represent the decision vector of the least squares problem, in this case the initial Cartesian state:
\begin{equation}
    \mathbf{x} =
    \begin{bmatrix}
        \mathbf{r}_0 \\
        \mathbf{v}_0
    \end{bmatrix} \text{,}
\end{equation}
where $\mathbf{r}_0$ and $\mathbf{v}_0$ are the initial position and state vectors respectively. In this case, the first pseudo-observation is used to initialise the estimator as it is expected to be sufficiently close to the optimal solution. \ac{SRP} can be estimated by including the corresponding parameter in the decision vector:
\begin{equation}
    \mathbf{x} =
    \begin{bmatrix}
        \mathbf{r}_0 \\
        \mathbf{v}_0 \\
        C_R
    \end{bmatrix} \text{,}
\end{equation}
where $C_R$ is the reflection coefficient.

The estimation process is expressed as an optimisation problem with the objective of minimising the sum of the square of the residuals:
\begin{align}
    \min_{\mathbf{x}} J(\mathbf{x}) &= \frac{1}{2} \mathbf{b}^T \mathbf{W} \mathbf{b} \text{,} \\
                                    &= \frac{1}{2} \left( \mathbf{y} - \mathbf{A} \mathbf{x} \right)^T \mathbf{W} \left( \mathbf{y} - \mathbf{A} \mathbf{x} \right) \text{,}
\end{align}
where $J(\mathbf{x})$ is the objective function, and $\mathbf{W}$ is the weighting matrix, a diagonal matrix containing weights to account for the expected observation noise:
\begin{equation}
    \mathbf{W} = \begin{bmatrix}
        \frac{1}{\sigma_1} \\
        & \frac{1}{\sigma_2} \\
        & & \ddots \\
        & & & \frac{1}{\sigma_n}
    \end{bmatrix} \text{,}
\end{equation}
where $\sigma_i$ is the standard deviation of the measurement noise for the $i$-th observation variable.

It can be shown that the minimisation problem can be expressed as the solution to a linear system~\citep{vallado_fundamentals_2013,tapley_statistical_2004}, known as the Gauss-Newton method:
\begin{equation}
    \left( \mathbf{A}_i^T \mathbf{W} \mathbf{A}_i \right) \delta \mathbf{x}_i = \mathbf{A}_i^T \mathbf{W} \mathbf{b}_i \text{,}
\end{equation}
\begin{equation}
    \mathbf{x}_{i+1} = \mathbf{x}_i + \delta \mathbf{x}_i \text{,}
\end{equation}
where $\delta \mathbf{x}_i$ is the best estimate of the optimal decision vector update for the $i$-th iteration. Due to the non-linearity of the dynamical system, the minimisation is iterated by re-linearising at each solution. This iteration continues until convergence is reached.

For this study, the Levenberg-Marquardt method~\citep{levenberg_method_1944,marquardt_algorithm_1963} was used instead. This enhancement of the Gauss-Newton method adaptively updates the solution by introducing a gradient descent term into the update equation:
\begin{equation}
    \left[ \mathbf{A}_i^T \mathbf{W} \mathbf{A}_i + \lambda\ \text{diag}\left( \mathbf{A}_i^T \mathbf{W} \mathbf{A}_i \right) \right] \delta \mathbf{x}_i = \mathbf{A}_i^T \mathbf{W} \mathbf{b}_i \text{,} \label{eq:LM_iter}
\end{equation}
where $\lambda$ is the damping coefficient, which determines the importance of the gradient descent update (the second term in brackets in \cref{eq:LM_iter}) relative to the Gauss-Newton update. The method initially prioritises gradient descent. The damping coefficient is reduced during iteration to increase the influence of the Gauss-Newton solution as the solver approaches the local minimum.

\subsection{Two-line Element Set Sampling}\label{s:tle_set_sampling}

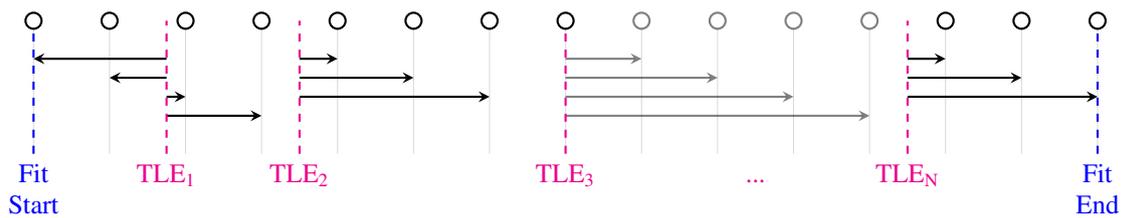
\begin{figure}[htb]
    \centering

    \vspace{0.5\baselineskip}

    \begin{tikzpicture}
        \draw[blue, dashed, thick] (0,-1.75) node[below, align=center]{Fit \\ Start} -- +(0, 1.75);
        \draw[blue, dashed, thick] (14,-1.75) node[below, align=center]{Fit \\ End} -- +(0, 1.75);

        \foreach \x in {0,...,14}
            \draw[black, opacity=0.15] (\x, -1.75) -- (\x, 0);

        \draw[black, -stealth, thick] (1.75, -0.5)  -- +(-1.75, 0);
        \draw[black, -stealth, thick] (1.75, -0.75) -- +(-0.75, 0);
        \draw[black, -stealth, thick] (1.75, -1.0)  -- +(0.25, 0);
        \draw[black, -stealth, thick] (1.75, -1.25) -- +(1.25, 0);

        \draw[black, -stealth, thick] (3.5, -0.5)  -- +(0.5, 0);
        \draw[black, -stealth, thick] (3.5, -0.75) -- +(1.5, 0);
        \draw[black, -stealth, thick] (3.5, -1.0)  -- +(2.5, 0);

        \draw[black, -stealth, thick, opacity=0.5] (7, -0.5)  -- +(1, 0);
        \draw[black, -stealth, thick, opacity=0.5] (7, -0.75) -- +(2, 0);
        \draw[black, -stealth, thick, opacity=0.5] (7, -1.0)  -- +(3, 0);
        \draw[black, -stealth, thick, opacity=0.5] (7, -1.25) -- +(4, 0);

        \draw[black, -stealth, thick] (11.5, -0.5)  -- +(0.5, 0);
        \draw[black, -stealth, thick] (11.5, -0.75) -- +(1.5, 0);
        \draw[black, -stealth, thick] (11.5, -1.0)  -- +(2.5, 0);

        \draw[magenta, dashed, thick] (1.75, -1.75) node[below] {TLE\textsubscript{1}} -- +(0, 1.75);
        \draw[magenta, dashed, thick] (3.5, -1.75)  node[below] {TLE\textsubscript{2}} -- +(0, 1.75);
        \draw[magenta, dashed, thick] (7, -1.75)    node[below] {TLE\textsubscript{3}} -- +(0, 1.75);
        \draw[magenta,  thick, thick] (9.5, -1.75)  node[below] {\vphantom{TLE\textsubscript{4}}...};
        \draw[magenta, dashed, thick] (11.5, -1.75) node[below] {TLE\textsubscript{N}} -- +(0, 1.75);

        \foreach \x in {0,...,7}
            \draw[fill=white, thick] (\x,0) circle (3pt);
        \foreach \x in {8,...,11}
            \fill[white, thick] (\x,0) circle (3pt);
        \foreach \x in {8,...,11}
            \draw[fill=white, thick, opacity=0.5] (\x,0) circle (3pt);
        \foreach \x in {12,...,14}
            \draw[fill=white, thick] (\x,0) circle (3pt);
    \end{tikzpicture}

    \caption{\acs{TLE} sampling strategy.}
    \label{f:tle_sampling}
\end{figure}

The sampling strategy proposed by \cite{levit_improved_2011} was used to generate pseudo-observations. Sampling times were equally spaced between the start and end epochs of the fit window, as illustrated in \cref{f:tle_sampling}. The state at given time was calculated by propagating the most recent preceding \ac{TLE} forward in time with SGP4. For sample times before the first \ac{TLE}, states were calculated by back-propagating.

For the satellites considered in this analysis, \acp{TLE} are released approximately every one or two days. Nevertheless, this frequency can vary and, additionally, in some cases \acp{TLE} can be re-issued to update parameters. This can create cases were, for a given epoch, multiple \acp{TLE} exist for a single object. To avoid this ambiguity, the most recently released \ac{TLE} is used as it represents the most up-to-date information.

Sharp discontinuities in object state can be introduced when transitioning between \acp{TLE} due to the updated orbital elements~\citep{araya_spectral_nodate}. Nevertheless, for many objects in \ac{MEO} and \ac{GEO}, the dynamical model is sufficiently accurate that the size of these discontinuities is relatively small. Under the current implementation, no smoothing between \acp{TLE} is conducted as it is not expected to have a significant impact on fit quality.

\cite{vallado_improved_2013} noted that \enquote{TLEs possess potentially wide variability} in quality. For a fit on a single poor quality \ac{TLE}, this can have a significant impact on the performance of a fit with respect to the truth. However, the influence of a single \enquote{bad} \ac{TLE} can be somewhat mitigated through the inclusion of multiple \acp{TLE} in the fit window~\citep{vallado_improved_2013}.

The weighting matrix was defined using \ac{TLE} noise estimates from literature, as presented in \cref{t:20231113_geul} in the \ac{RTN} frame. The weights are defined by standard deviations in inertial space, therefore for each epoch, the uncertainties were rotated from the \ac{RTN} frame to the inertial frame of the corresponding pseudo-observation, considering only the diagonal terms when constructing the weighting matrix. It should be noted that these estimates were derived from objects in \ac{LEO} and, therefore, represent an underestimate for the noise at higher altitudes, where \ac{TLE} uncertainty is typically larger~\citep{racelis_high-integrity_2018}.

\begin{table}[htb]
    \centering

    \caption{Mean \acs{TLE} uncertainty in the \acs{RTN} frame at epoch, for \enquote{enhanced} \acsp{TLE} released after 2013. Adapted from~\cite{geul_tle_2017}.}
    \label{t:20231113_geul}

    \sisetup{
        exponent-mode=input,
        table-format=1.1e+1,
        retain-zero-exponent=true
    }

    \renewcommand{\arraystretch}{1.1}

    \begin{tabular}{l S S S}
        \toprule
        Type            & {${\sigma_{R}}$}  & {${\sigma_{T}}$}  & {${\sigma_{N}}$}  \\
        \midrule
        Position [m]    & 1.2e2             & 2.0e3             & 8.0e1             \\
        Velocity [m/s]  & 2.4e0             & 1.3e-1            & 6.8e-2            \\
        \bottomrule
    \end{tabular}
\end{table}

\subsection{Calibration Satellites}

Four calibration satellites were considered for initial testing: both pairs each of the \ac{LAGEOS} and Etalon satellites, as listed in \cref{t:calibration_satellites}. These calibration satellites were designed for geodesy and are covered in retro-reflectors which are designed to reflect beams of light back to their point of origin~\citep{pearlman_laser_2019}. These allow ground-based laser systems to make highly precise ranging measurements (typically at the centimetre level) based on the two-way time-of-flight of laser pulses.

\begin{table}[htb]
    \centering
    \caption{Approximate orbital elements of the calibration satellites (1\textsuperscript{st} January 2022).}
    \label{t:calibration_satellites}
    \begin{tabular}{l c c c c c}
    \toprule
    Name        & \acs{NORAD} ID & Semi-major axis [km] & Eccentricity [-] & Inclination [deg] \\
    \midrule
    LAGEOS-1    & 8820  & 12270 & 0.005 & 110  \\
    LAGEOS-2    & 22195 & 12160 & 0.014 & 52.6 \\
    Etalon 1    & 19751 & 25500 & 0.002 & 64.3 \\
    Etalon 2    & 20026 & 25500 & 0.002 & 65.6 \\
    \bottomrule
    \end{tabular}
\end{table}

These calibration satellites were selected due to a number of advantages: their altitudes (in \ac{MEO}) which reduced the influence of certain perturbations, such as atmospheric drag which is effectively not present; their high mass-to-area ratios which limited the effect of \ac{SRP}; and, most importantly, their role as geodesy satellites which meant that accurate ephemerides were readily available for validating fit accuracy. The lack of atmospheric drag and \ac{SRP} meant that these perturbations could be ignored. This simplified the numerical model and removed the need to estimate drag and reflection coefficients.

\enquote{Ground truth} data for the calibration satellites was provided by the \ac{ILRS}. Post-processed, high-precision ephemerides derived from \ac{SLR} were retrieved from \acs{NASA}'s \ac{CDDIS}\footnote{Available at: \url{https://cddis.nasa.gov/archive/slr/products/orbits/} (accessed on 28/10/2024)} and \acs{ESA}'s Navigation Office\footnote{Available at: \url{http://navigation-office.esa.int/products/slr-products/} (accessed on 28/10/2024)}.

\subsection{Test Satellites}

Further testing considered additional objects, as listed in \cref{t:test_satellites}. These include Navstar~1 and Navstar~2, the first two \ac{GPS} satellites; and Astra~1E and Astra~1H, two telecommunications satellites. These provide examples in \ac{MEO} and \ac{GEO}/\ac{GSO}, respectively with the Navstar satellites having similar orbits to the Etalon satellites. Importantly, examination of the \ac{TLE} series for each of these objects showed no evidence of manoeuvres, meaning that the satellites could be considered passive, allowing the study to focus on perturbation modelling in the \acp{TLE} themselves.

\begin{table}[htb]
    \centering
    \caption{Approximate orbital elements of the test satellites (1\textsuperscript{st} January 2022).}
    \label{t:test_satellites}
    \begin{tabular}{l c c c c c}
    \toprule
    Name        & \acs{NORAD} ID & Semi-major axis [km] & Eccentricity [-] & Inclination [deg] \\
    \midrule
    Navstar~1   & 19751 & 26800 & 0.008 & 63.3 \\
    Navstar~2   & 20026 & 26500 & 0.050 & 63.8 \\
    Astra~1E    & 23686 & 42600 & 0.001 & 8.40 \\
    Astra~1H    & 25785 & 42500 & 0.001 & 7.12 \\
    \bottomrule
    \end{tabular}
\end{table}

Unlike the calibration satellites, precise ephemerides, either from \ac{SLR} or pseudo-ranging, are not available for the test satellites. This means that \acs{TLE}-based errors metrics are required to evaluation fit quality.

\subsection{Physical Model}

Computational performance was an important consideration as it is planned to extend this estimation process to the entire \ac{TLE} catalogue. It was decided to use a \enquote{minimal model} where model parameters were tuned to minimise computational cost without significantly impacting the accuracy of the estimation process.

During initial testing of the \ac{P-OD} process, gravitational perturbations were found to be the most critical, therefore a geopotential model to account for the non-sphericity of the Earth, and third body perturbations resulting from the Moon and the Sun were included in the physical model.

A convergence study was used to evaluate the influence of geopotential degree and order on propagations, using a high degree and order model (50-by-50) as a reference. It was found that a truncated 10-by-10 model was sufficient to achieve less than \SI{40}{\metre} position error after 30~days of propagation with respect to the 50-by-50 geopotential model, for the worst case which corresponds to the altitudes of \acs{LAGEOS}-1 and -2. This was considered adequate for this analysis. Similarly, solid tides were found to have a minimal effect.

The influence of \ac{SRP} was evaluated by comparing propagations with various reflection coefficients against a propagation without \ac{SRP}. In the case of \acs{LAGEOS}-1 and -2, an assumed $C_R A / m$ of \SI{1.0e-3}{\metre^2/\kilogram} resulted in a position difference after 30~days of less than \SI{30}{\metre} and, for Etalon 1 and 2, an assumed $C_R A / m$ of \SI{1.3e-3}{\metre^2/\kilogram} resulted in a position difference after 30~days of less than \SI{150}{\metre}. Consequently, the influence of \ac{SRP} could be ignored for the calibration satellites. \ac{SRP} was included and estimated for the test satellite fits, however it should be noted that the observability of \ac{SRP} was limited: position errors due to \ac{SRP} had similar magnitudes to the noise in the \acp{TLE}, in the order of \SI{10}{\kilo\metre} after 30~days for an assumed $C_R A / m$ of \SI{3.0e-2}{\metre^2/\kilogram}.

Propagation was performed using the THALASSA propagator \citep{amato_thalassa_2018}. A 10-by-10 truncated form of the GRIM5-S1 gravity field model \citep{biancale_new_2000} was used for geopotential modelling. \acs{JPL}'s DE431 planetary and lunar ephemerides \citep{folkner_planetary_nodate} were used to generate state vectors for the Sun and Moon when calculating lunisolar perturbations. A \enquote{cannonball} model was used for \ac{SRP} with a conical Earth shadow model.

The EDromo(c) propagation method~\citep{bau_edromo_2014} was used in THALASSA, providing the benefits of regularised formulations, in this case primarily improved computational performance. Numerical integration is provided by the \ac{LSODAR}. This solver switches automatically between non-stiff and stiff methods, Adams and BDF respectively, and includes root finding capabilities~\citep{hindmarsh_odepack_1982}. A tolerance of \num{1e-10} was found to provide a good balance between computational performance and integration error. 

\subsection{Fit Quality Evaluation}

Prediction accuracy was evaluated primarily through the post-fit position \ac{RMSE}:
\begin{equation}
    \Delta r_{\text{RMSE}} = \sqrt{\frac{1}{N} \sum_i^N || \mathbf{r}_{i}^* - \mathbf{r}_i ||^2 } \text{,}
\end{equation}
where $N$ is the number of samples; and $\mathbf{r}_{i}^*$ and $\mathbf{r}_i$ are the predicted and true positions, respectively, at the $i$-th epoch. An alternative formulation of this metric, for use in the absence of \ac{SLR} data, replaces the true positions with those derived from future \acp{TLE}, as discussed in \cref{s:error_metric_correlation}.

The uncertainty of a fit can be calculated directly during the estimation process and represented by a covariance matrix:
\begin{equation}
    \mathbf{P} = \left( \mathbf{A}^T \mathbf{W} \mathbf{A} \right)^{-1} \text{,}
\end{equation}
however, this requires a well-tuned weighting matrix, therefore the sample covariance matrix was used instead, calculated with the fit residuals in the \ac{RTN} frame.

Covariances estimated during the \ac{P-OD} process can be compared by considering the size of the region of uncertainty. Assuming the position to be described by a Gaussian distribution, its probability density is equal to the $3\sigma$ value on the ellipsoid of volume:
\begin{equation}
    V_{3\sigma} = \frac{4}{3} \pi \left(3\sigma_{r,1}\right) \left(3\sigma_{r,2}\right) \left(3\sigma_{r,3}\right) = 36 \pi \left(\sigma_{r,1}\right) \left(\sigma_{r,2}\right) \left(\sigma_{r,3}\right) \text{,}
\end{equation}
where $\sigma_{r,1}$, $\sigma_{r,2}$, and $\sigma_{r,3}$ are the position standard deviations in the principal axes, i.e., the eigenvalues of the position covariance matrix. As an uncertainty metric, we consider the equivalent radius of the uncertainty ellipsoid, that is the radius of a sphere with the same volume as the 3-$\sigma$ ellipsoid:
\begin{equation}
    R_{3\sigma} = \sqrt[3]{\left(3\sigma_{r,1}\right) \left(3\sigma_{r,2}\right) \left(3\sigma_{r,3}\right)} = 3 \sqrt[3]{\left(\sigma_{r,1}\right) \left(\sigma_{r,2}\right) \left(\sigma_{r,3}\right)} \text{.}
\end{equation}

\subsection{Two-line Element Set Debiasing}\label{s:debiasingmethod}

Systematic biases in the \acp{TLE} cause significant variations in fit quality throughout the year. \cite{ly_correcting_2020} highlighted seasonal and periodic variations in \ac{TLE} quality, primarily affecting the along-track component of the satellite's position. These variation were correlated with the cycles of the Sun and Moon, suggesting mismodelling of lunisolar perturbations in SGP4.

To demonstrate the concept of \acs{TLE} debiasing, a simplified sinusoidal model was proposed for demonstrating the estimation of along-track error~\citep{hallgarten_la_casta_revisiting_2024}:
\begin{equation}\label{e:bias_model}
    \Delta \theta_T (t) \approx a \sin\left( \frac{2 \pi}{b} \left(t + c\right) \right) + d \text{,}
\end{equation}
where $\Delta \theta_T = \Delta{r}_T/r$ is the along-track angular error; $a$, $b$, $c$, $d$ are the bias amplitude, period, phase offset, and mean offset respectively; and $t$ is the time, in this case defined as an offset from a given epoch. The \ac{RTN} frame used for the bias model is defined by the state of the \ac{TLE} so that estimations from the model apply relative to the \ac{TLE} itself.

\begin{figure}[p]
    \centering

    \includegraphics{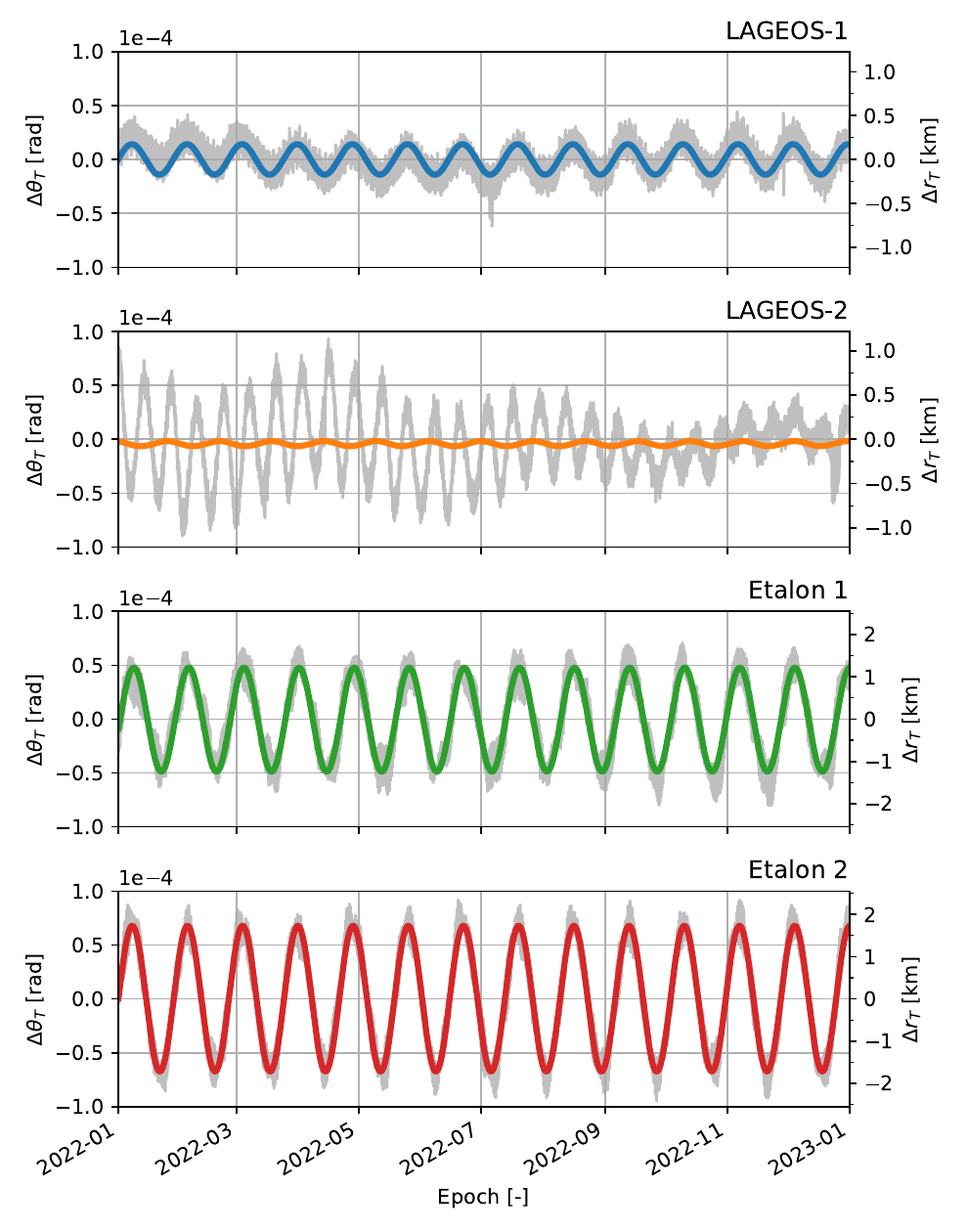}

    \caption{Along-track errors of \acsp{TLE} with respect to \acs{SLR}. The \acs{SLR} data and fitted models are presented in grey and colour respectively.}

    \label{f:transverse_errors_plot}
\end{figure}

The fitted parameters for each of the calibration satellites, estimated using non-linear batch least squares, are presented in \cref{t:transverse_errors_model_parameters} and the corresponding models illustrated in \cref{f:transverse_errors_plot}. The Etalon~1 and~2 satellites showed good agreement with the model, with their respective parameters; however, \acs{LAGEOS}-1 showed slightly different behaviour with an additional longer term variation, while \acs{LAGEOS}-2 showed a complete lack of agreement with the simplified model caused by the presence of error oscillations at higher frequency.

\begin{table}[htb]
    \centering

    \caption{Fitted parameters for the bias model, expressing time in days since midnight on the 1\textsuperscript{st} January 2022.}
    \label{t:transverse_errors_model_parameters}

    \sisetup{exponent-mode = input}

    \begin{tabular}{l S[table-format=1.2e+1] S[table-format=2.1] S[table-format=-1.2e+1] S[table-format=-1.2e+1]}
    \toprule
    Satellite       & {$a$ [rad]}   & {$b$ [days]}  & {$c$ [days]}  & {$d$ [rad]}   \\
    \midrule
    LAGEOS-1        & 1.41e-5       & 27.5          & 6.82e-2       & -1.79e-8      \\
    LAGEOS-2        & 2.28e-6       & 26.6          & -1.19e1       & -4.00e-6      \\
    Etalon 1        & 4.82e-5       & 27.5          & -9.17e-1      & -4.43e-7      \\
    Etalon 2        & 6.75e-5       & 27.6          & -1.80e-1      & 7.79e-8       \\
    \bottomrule
    \end{tabular}
\end{table}

The period of the along-track error model had significant agreement between \acs{LAGEOS}-1 and Etalon~1 and~2 at approximately 27.5~days. This is very close to the Moon's sidereal orbital period (approximately 27.3~days), supporting the suggestion that mismodelling of the Moon's third-body perturbation in SGP4 is a contributing factor to the along-track error. The mismodelling error may result from either, or a combination of, the lunar ephemerides used by SGP4 or a lack of modelling of lunar long-periodic terms. The phase offsets also showed good agreement, varying by less than one day between the satellites, suggesting that it may be possible to constrain these parameters to constants across all of the satellites.

Pseudo-observations are debiased by subtracting the modelled biases from the observation vector:
\begin{equation}
    \mathbf{y}_\text{debiased}  = \mathbf{y} - \mathbf{y}_\text{bias} \text{,}
\end{equation}
resulting in an updated version of \cref{e:residual_noa} which considers the effect of the bias:
\begin{align}
    \mathbf{b}_\text{debiased} &= \mathbf{y}_\text{debiased} - \mathbf{y}^* \text{,} \\
    &= \mathbf{y} - \mathbf{y}_\text{bias} - \mathbf{y}^* \text{,}
\end{align}
where $\mathbf{b}_\text{debiased}$ is the debiased residual vector.

\section{Bias Impact on P-OD Prediction Accuracy}\label{s:results}

\subsection{Calibration Satellite Fits}\label{s:calibration_fits}

\begin{figure}[p]
    \centering
    \includegraphics{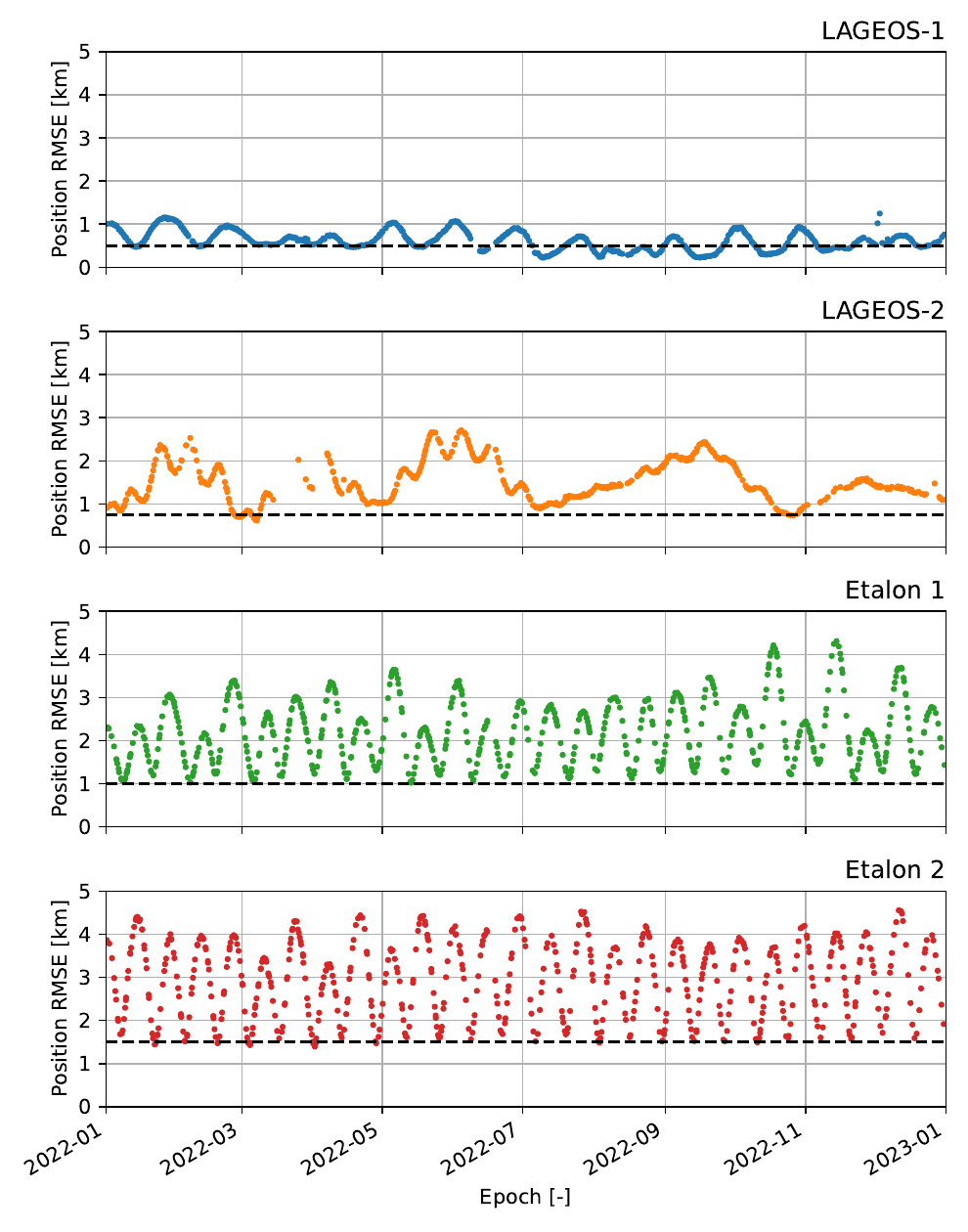}
    \caption{Position \acs{RMSE}, with respect to \acs{SLR}, during 30 days post-epoch for calibration satellite \acsp{TLE}. The dashed lines represent the established baselines for each satellite.}
    \label{f:tle_rmse_cal}
\end{figure}

The quality of the calibration satellites' \acp{TLE} was evaluated fits to provide a benchmark for fit performance. The position \acp{RMSE} for the \acp{TLE} of the calibration satellites are presented in \cref{f:tle_rmse_cal}, calculated by propagating each of the \acp{TLE} for 30~days after their epoch and evaluating their \acp{RMSE} with respect to the \acs{SLR}-derived states. All four calibration satellites showed periodic variations in \ac{RMSE} throughout the year, albeit with differing behaviours. \acs{LAGEOS}-1 had largely periodic behaviour which remained below \SI{1}{\kilo\metre} for the majority of the year. Etalon 1 and 2 had higher \acp{RMSE} with larger oscillations which remained bounded between approximately \SIrange{1}{4.5}{\kilo\metre} and \SIrange{1.5}{4.5}{\kilo\metre} respectively. Notably, the frequency of the error behaviour for Etalon~1 and~2 was approximately double that of \acs{LAGEOS}-1. \acs{LAGEOS}-2 had the most irregular behaviour with no obvious pattern, remaining bounded between approximately \SIrange{0.75}{2.5}{\kilo\metre}. The lower bounds of these errors provided a useful point of reference: consistently meeting the bound throughout the year alone would be a significant improvement as this would lead to the removal of the oscillations and more consistent state estimates.

\begin{figure}[htbp]
    \centering

    \begin{subfigure}[c]{0.5\textwidth}
        \centering
        \includegraphics{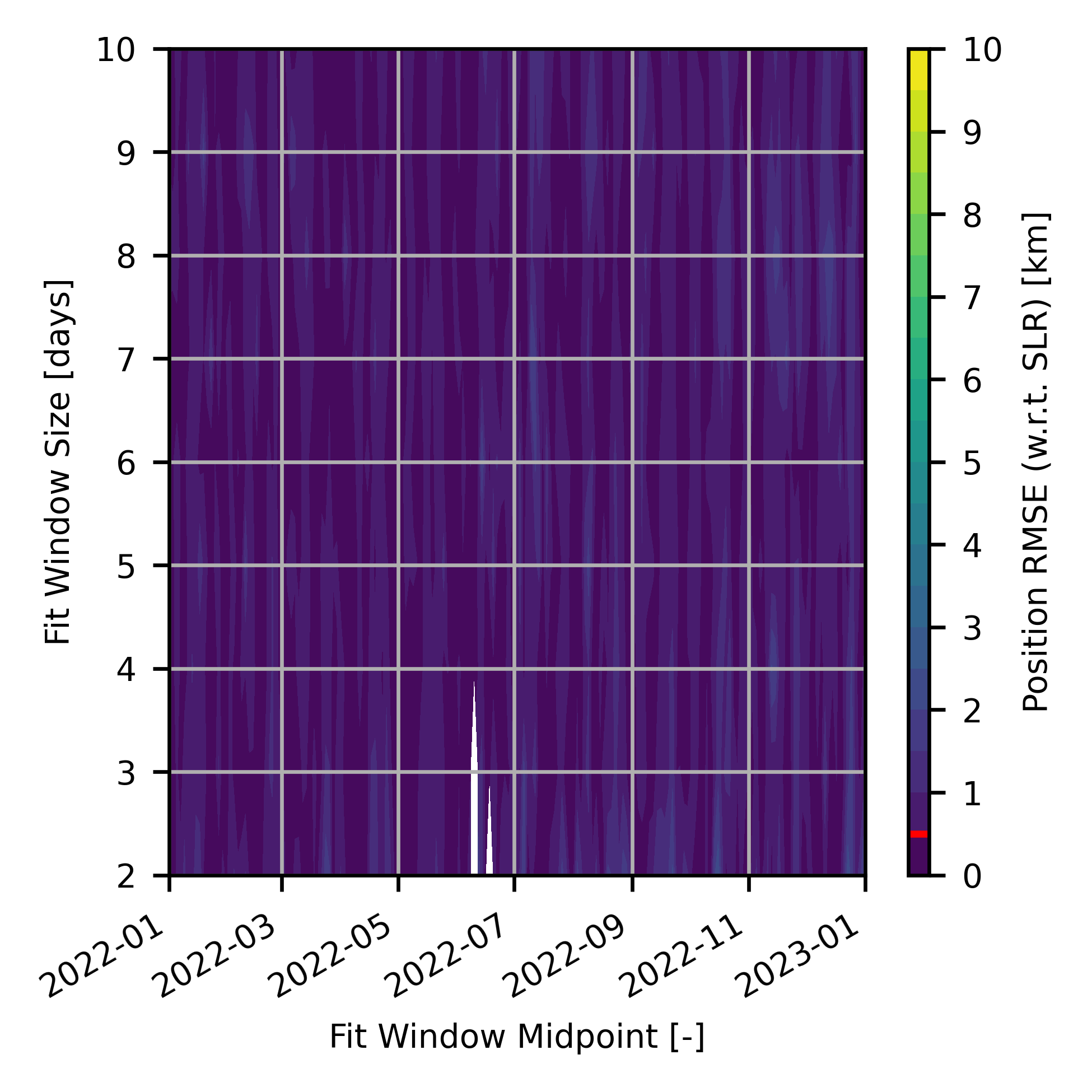}
        \caption{\acs{LAGEOS}-1.}
    \end{subfigure}%
    \begin{subfigure}[c]{0.5\textwidth}
        \centering
        \includegraphics{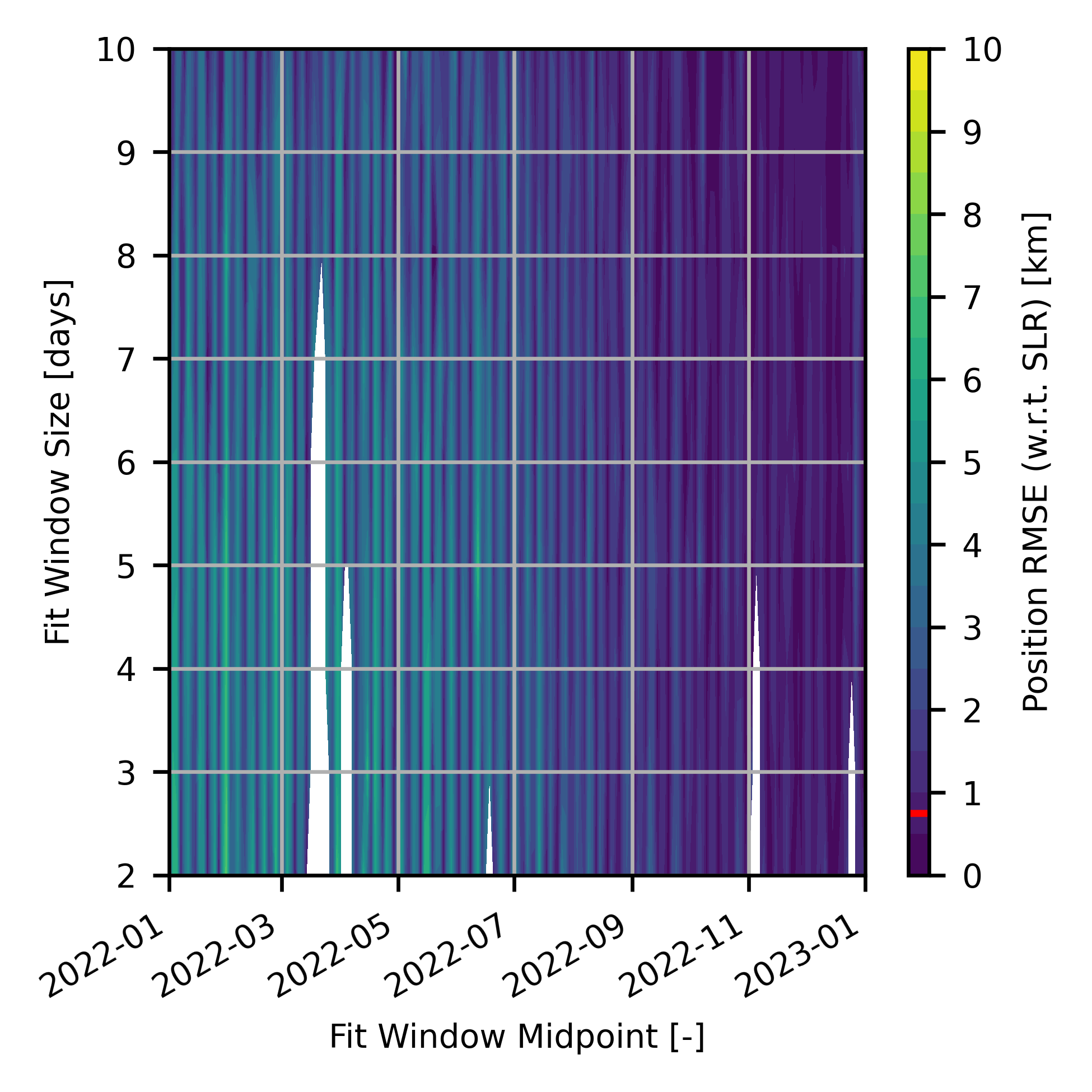}
        \caption{\acs{LAGEOS}-2.}
    \end{subfigure}

    \vspace{0.5\baselineskip}

    \begin{subfigure}[c]{0.5\textwidth}
        \centering
        \includegraphics{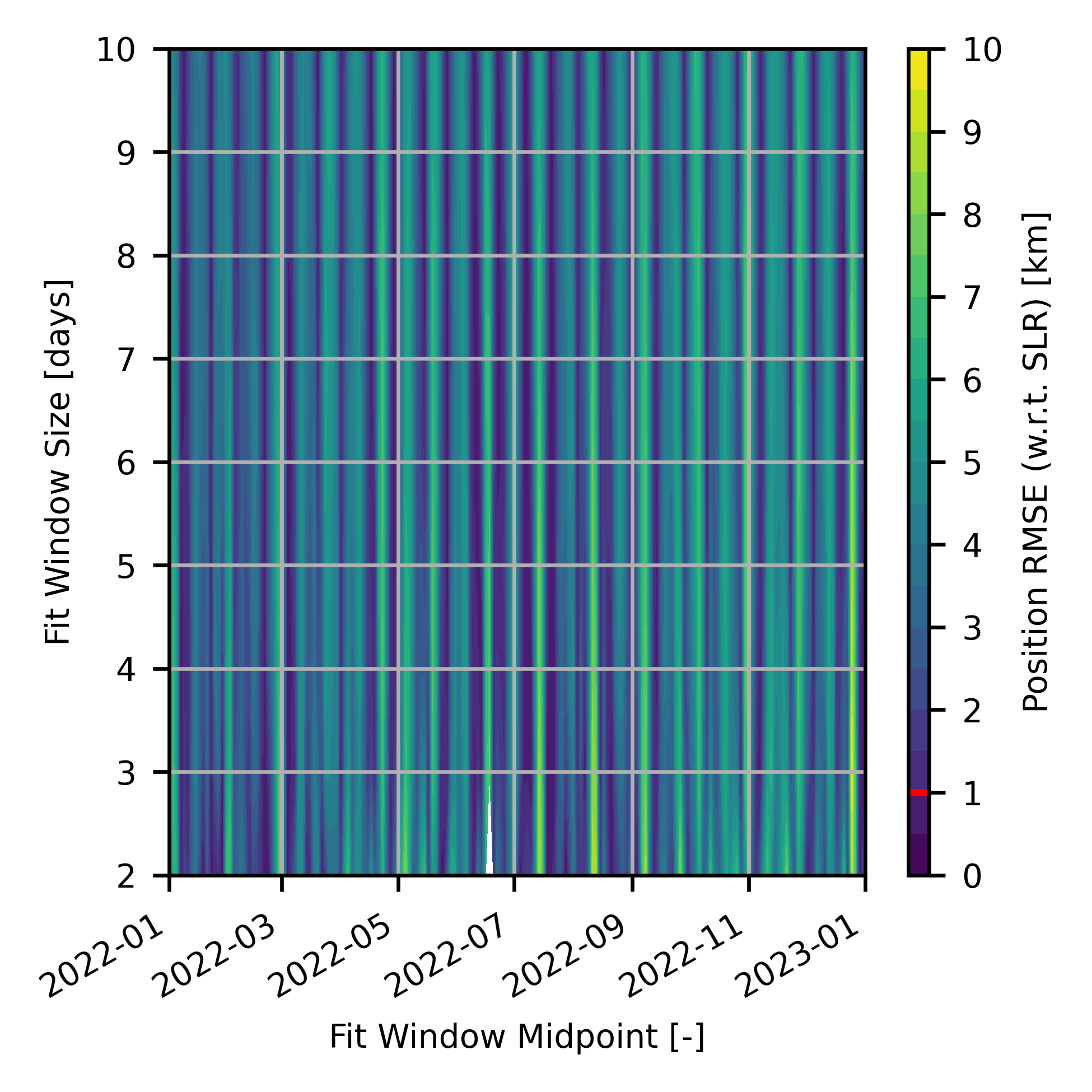}
        \caption{Etalon~1.}
    \end{subfigure}%
    \begin{subfigure}[c]{0.5\textwidth}
        \centering
        \includegraphics{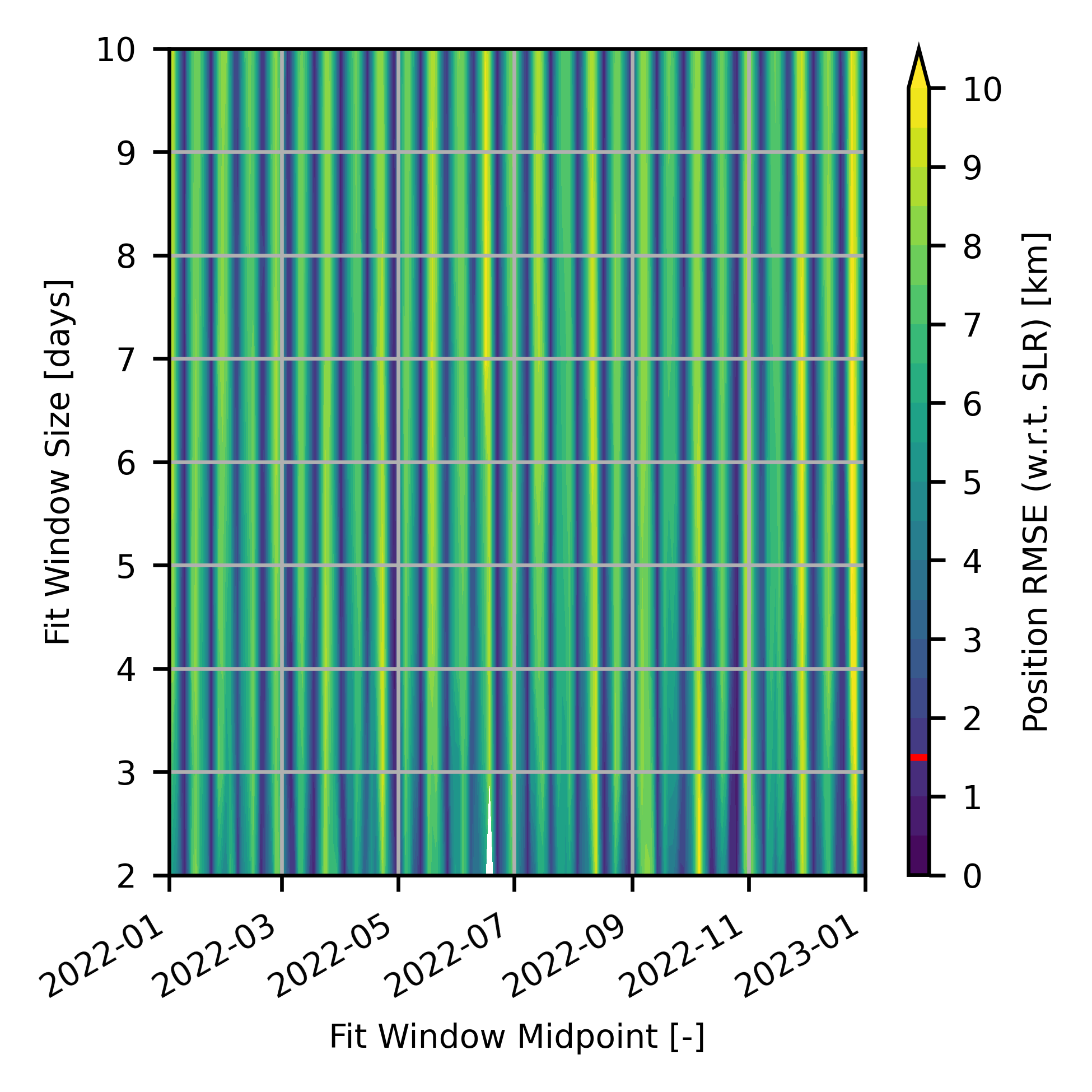}
        \caption{Etalon~2.}
    \end{subfigure}

    \vspace{0.25\baselineskip}
    
    \caption{Position \acs{RMSE} during the 30~days post-fit for short-term biased fits. The red marker in the colour bar indicates the \acs{RMSE} baseline derived from \acsp{TLE}. Note that white regions indicate failed cases due to an insufficient number of \acsp{TLE}.}
    \label{f:cal_short_biased}
\end{figure}

\begin{figure}[htbp]
    \centering

    \begin{subfigure}[c]{0.5\textwidth}
        \centering
        \includegraphics{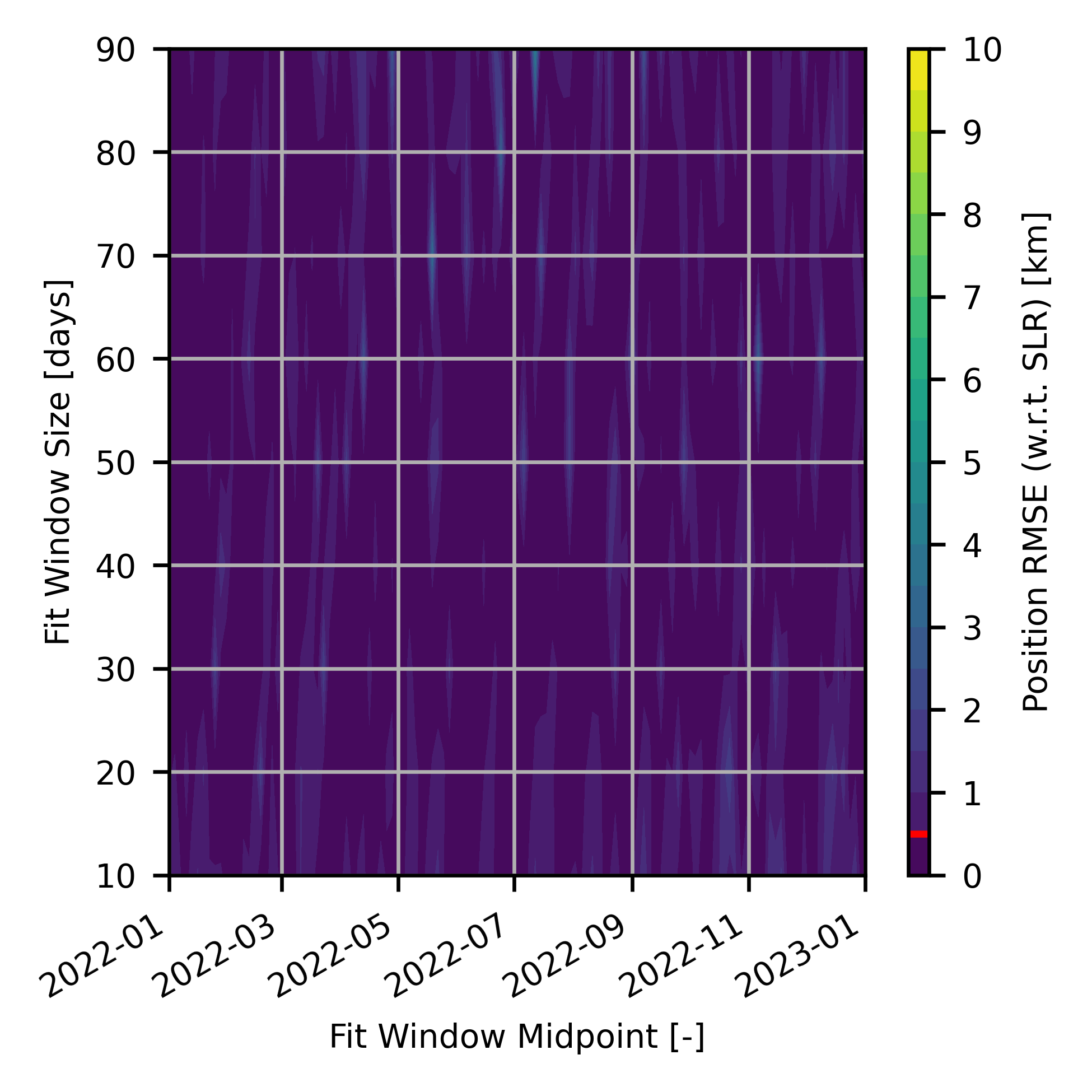}
        \caption{\acs{LAGEOS}-1.}
    \end{subfigure}%
    \begin{subfigure}[c]{0.5\textwidth}
        \centering
        \includegraphics{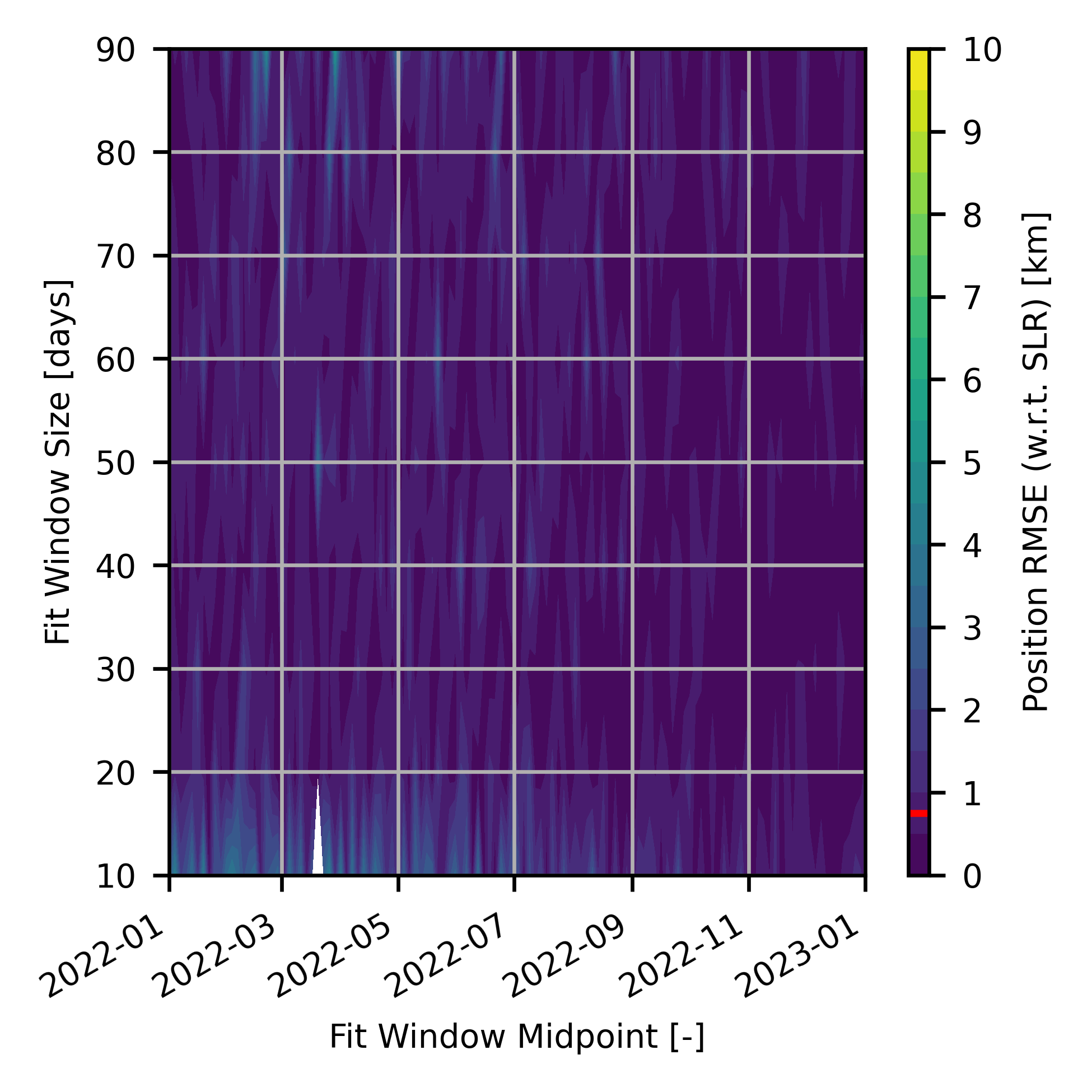}
        \caption{\acs{LAGEOS}-2.}
    \end{subfigure}

    \vspace{0.5\baselineskip}

    \begin{subfigure}[c]{0.5\textwidth}
        \centering
        \includegraphics{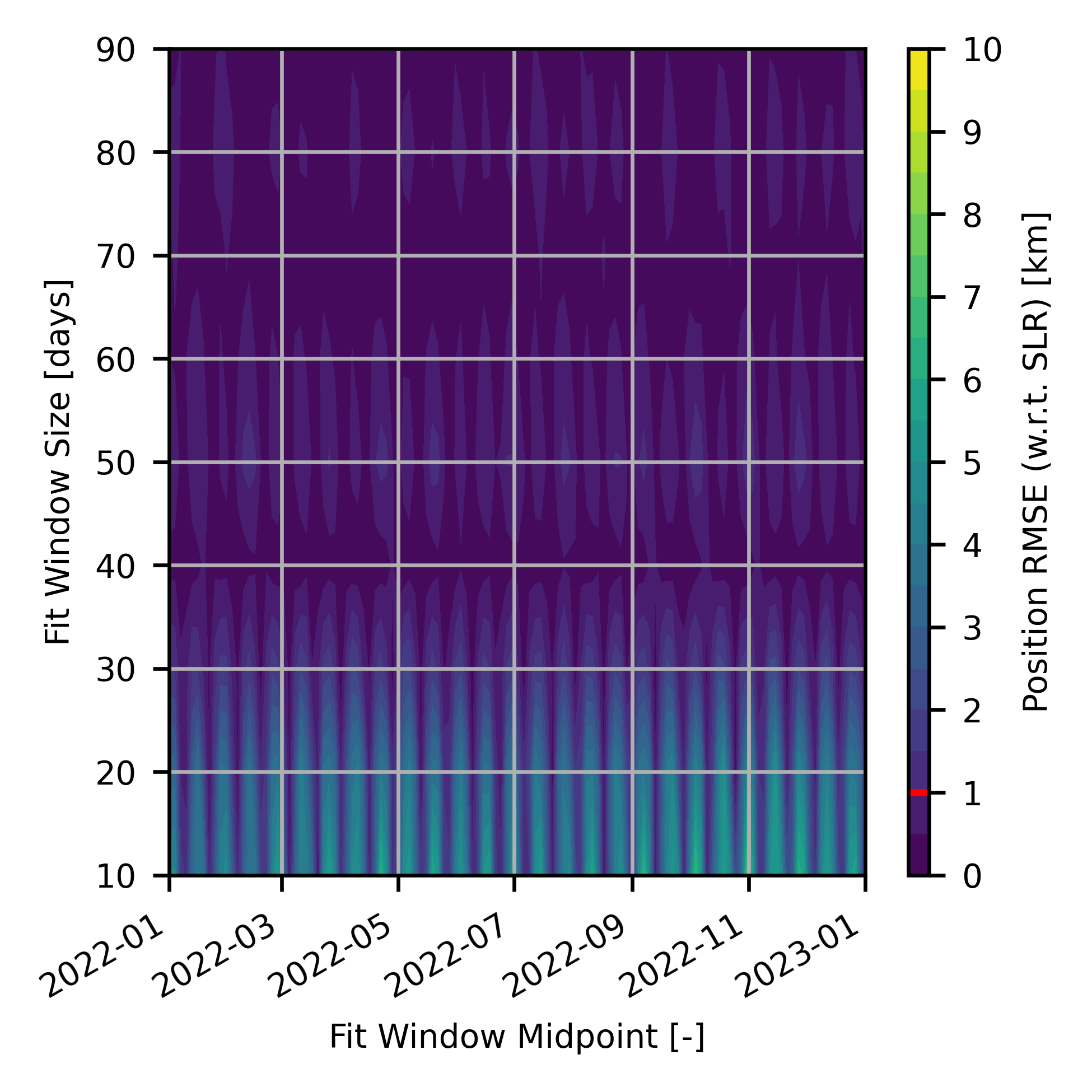}
        \caption{Etalon~1.}
    \end{subfigure}%
    \begin{subfigure}[c]{0.5\textwidth}
        \centering
        \includegraphics{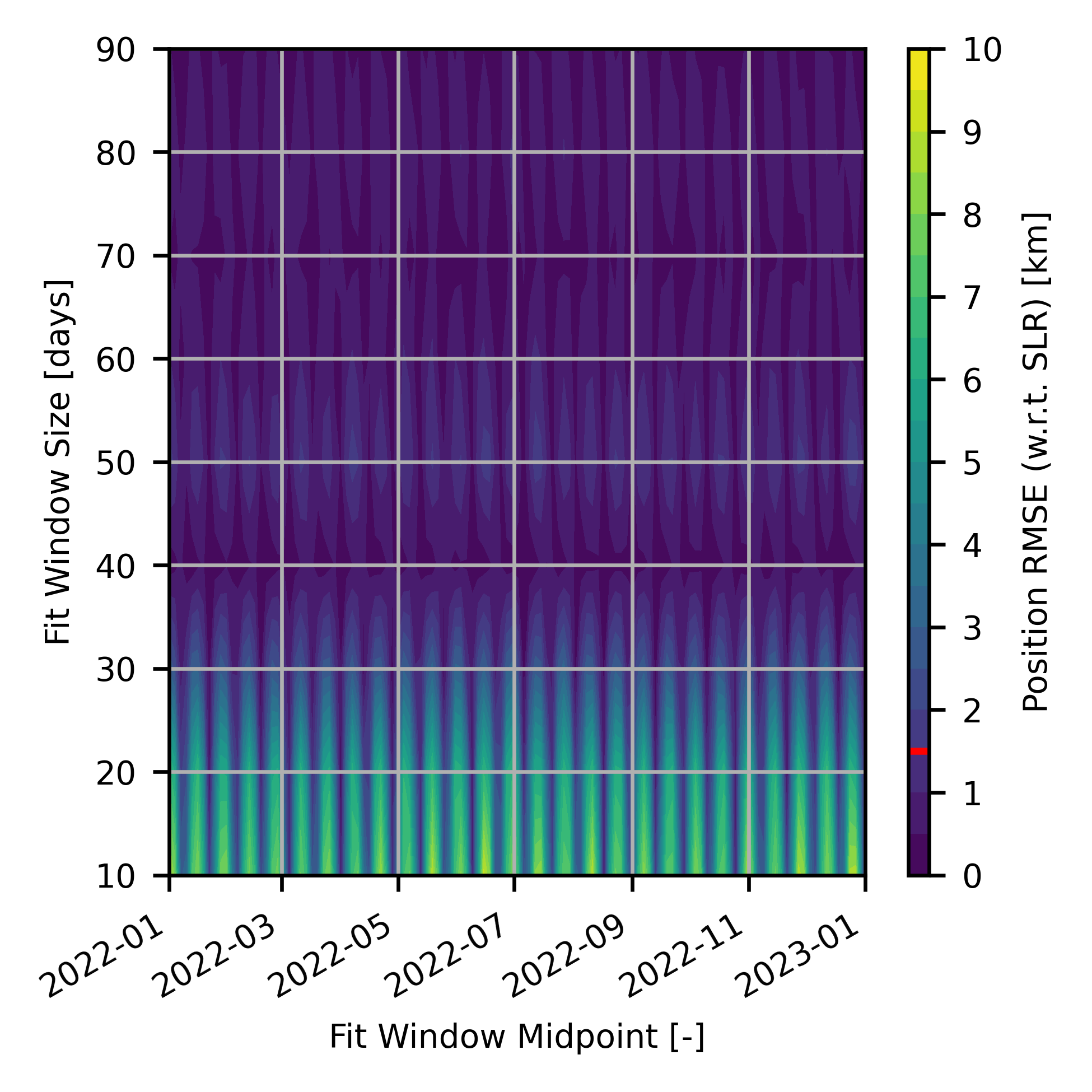}
        \caption{Etalon~2.}
    \end{subfigure}

    \vspace{0.25\baselineskip}
    
    \caption{Position \acs{RMSE} during the 30~days post-fit for long-term biased fits. The red marker in the colour bar indicates the \acs{RMSE} baseline derived from \acsp{TLE}. Note that white regions indicate failed cases due to an insufficient number of \acsp{TLE}.}
    \label{f:cal_long_biased}
\end{figure}

Fits on the calibration satellites were conducted using \ac{TLE} and \ac{SLR} data from the year 2022, considering combinations of fit midpoints and window sizes, as presented in \cref{t:fit_parameters}. The motivation of these tests was to investigate the influence of the fit epoch and fit window size on the quality of fits, and whether a specific combination of these parameters was particular effective in mitigating the impact of biases in pseudo-observations. The results of conducting these fits are presented in \cref{f:cal_short_biased,f:cal_long_biased}.

\begin{table}[htb]
    \centering

    \caption{Fit sweep parameters.}
    \label{t:fit_parameters}
    
    \begin{tabular}{l c c c}
    \toprule
    Term    & Fit midpoint period [days]    & Number of samples [-] & Fit window size [days]                \\
    \midrule
    Short   & 2                             & 100                   & 2, 3, 4, 5, 6, 7, 8, 9, 10            \\
    Long    & 3                             & 100                   & 10, 20, 30, 40, 50, 60, 70, 80, 90    \\
    \bottomrule
    \end{tabular}
\end{table}

Fits on all four of the calibration satellites showed similar behaviour for the short-term cases: fit error was primarily driven by the fit window midpoint, noticeable as vertical banding throughout the year, particularly for Etalon 1 and 2. Fit window size appeared to have minimal impact on the fit error for the short-term cases with increased sizes not being associated with any change in error. \acs{LAGEOS}-2 saw a significant improvement in fit quality towards the end of 2022, both in terms of typical error and the magnitude of the oscillations. This was associated with the reduction in along-track bias magnitude, as seen in \cref{f:transverse_errors_plot}.

For the long-term fits, the dependence on fit midpoint remained for smaller windows. Nevertheless, the variation throughout the year reduced as the size of the windows increased. Above 40~days, the oscillations in fit error became significantly smaller for all four satellites, eventually converging to errors significantly lower than those seen for the short-term fits. Notably, for Etalon~1 and~2, some slight banding reappeared for fit windows around 50~days and 80~days, which demonstrated that the periodic influence of the bias was not completely eliminated in all cases by extending the fit windows. In any case, for the extended windows, the post-fit errors were typically at or below the baselines established in \cref{f:tle_rmse_cal}, suggesting that, for long-term fits, prediction accuracy was improved over the best case \acs{TLE}-based predictions.

\begin{figure}[htb]
    \centering

    \begin{subfigure}[c]{0.5\textwidth}
        \centering
        \includegraphics{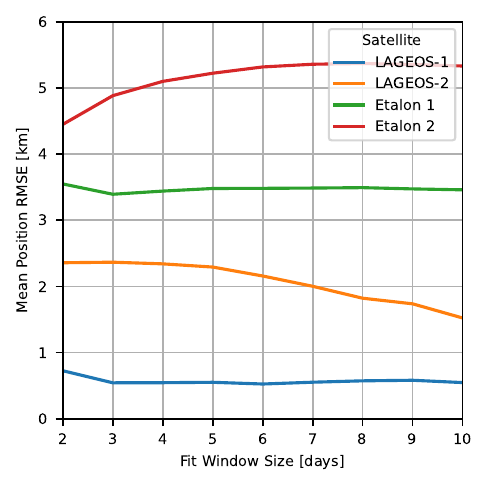}
        \caption{Short-term.}
    \end{subfigure}%
    \begin{subfigure}[c]{0.5\textwidth}
        \centering
        \includegraphics{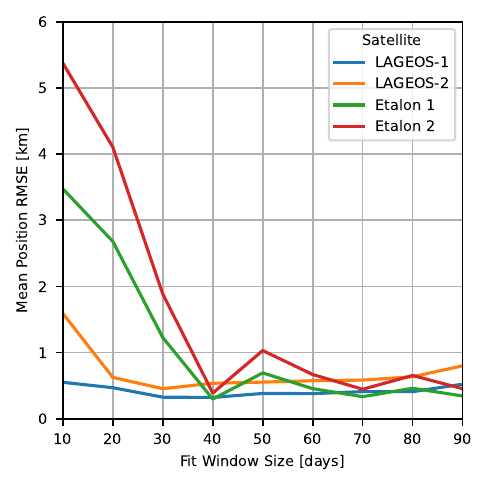}
        \caption{Long-term.}
    \end{subfigure}

    \vspace{0.25\baselineskip}
    
    \caption{Mean position \acs{RMSE} during the 30~days post-fit for biased fits.}
    \label{f:cal_mean_biased}
\end{figure}

The impact of increasing the size of the fit window was further illustrated by considering the aggregate behaviour on fit error for each object. The mean position \ac{RMSE} for each object as a function of the fit window size is presented in \cref{f:cal_mean_biased}. For the short-term fits, the mean error remained largely constant for all four objects, other than a slight reduction for \acs{LAGEOS}-2 and slight increase for Etalon~2. Significant reductions did not occur until the fit window size was increased past 10~days with all four objects reaching their respective baselines after 40~days. The slight banding for Etalon~1 and~2 at 50~days and 80~days was seen as slight increases in the mean error.

Importantly, many of the short-term fits had significantly higher errors than simply propagating \acp{TLE} directly with SGP4. For example, fits on Etalon~1 and~2 often had errors double that of the worst cases presented in \cref{f:tle_rmse_cal}. This indicated that \ac{P-OD} cannot be applied without careful selection of parameters nor without considering the impact of bias. In the absence of bias modelling, long fit windows are required to successfully apply \ac{P-OD} techniques to \ac{TLE} series.

\subsection{Error Metric Correlation}\label{s:error_metric_correlation}

\begin{figure}[htb]
    \centering

    \vspace{0.5\baselineskip}

    \begin{subfigure}[t]{0.45\textwidth}
        \centering

        \begin{tikzpicture}[scale=1.2]

        \draw[blue, thick, domain=0:4, samples=100, variable=\x] plot ({\x}, {sin (360 * \x)});
        \draw[red,  thick, domain=0:4, samples=100, variable=\x] plot ({\x}, {0.25 * sin (920 * \x)});

        \draw[black, thick, -stealth] (-0.25,0) -- (4.25, 0);
        
        \draw[blue, thick, stealth-stealth] (4.75, 0) -- (4.75, 1) node [midway, above, sloped] {TLE error};
        \draw[red,  thick, stealth-stealth] (5.25, 0) -- (5.25, 0.25) node [midway, above, sloped] {Fit error};

        \path (-0.5, -1.5) -- (5.5, 3.5);

        \end{tikzpicture}

        \caption{\acs{TLE} dominated, where the errors in the \acsp{TLE} themselves drive the error.}
    \end{subfigure}
    \hspace{0.075\textwidth}
    \begin{subfigure}[t]{0.45\textwidth}
        \centering

        \begin{tikzpicture}[scale=1.2]

        \draw[blue, thick, domain=0:4, samples=100, variable=\x] plot ({\x}, {sin (360 * \x)});
        \draw[red,  thick, domain=0:4, samples=100, variable=\x] plot ({\x}, {0.75 * \x + 0.25 * sin (920 * \x)});

        \draw[black, thick, -stealth] (-0.25,0) -- (4.25, 0);

        \draw[blue, thick, stealth-stealth] (4.75, 0) -- (4.75, 1) node [midway, above, sloped] {TLE error}; 
        \draw[red,  thick, stealth-stealth] (5.25, 0) -- (5.25, 3) node [midway, above, sloped] {Fit error}; 

        \path (-0.5, -1.5) -- (5.5, 3.5);

        \end{tikzpicture}

        \caption{Fit dominated, where the errors in the fit drive the error.}
    \end{subfigure}

    \vspace{0.25\baselineskip}
    
    \caption{Illustration of error metric scenarios.}
    \label{f:error_metric_demonstration}
\end{figure}
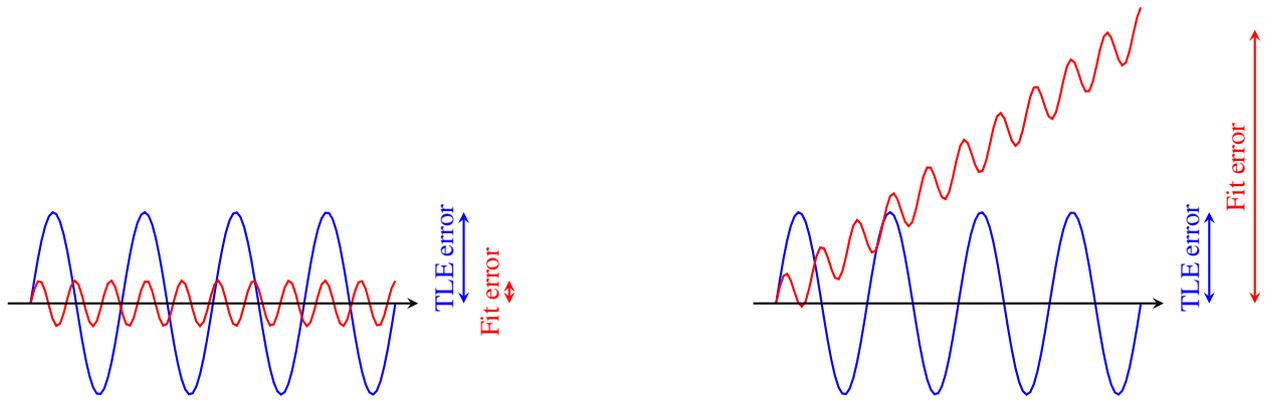

The \acs{TLE}-based error metric for estimating fit quality in the absence of \ac{SLR} data captures errors resulting from both the fit and from the bias and noise in the \acp{TLE} themselves. It was expected, therefore, that a lower bound would exist for this metric, representing the error of the \acp{TLE} alone for a hypothetical perfect fit with no error with respect to the true state of the satellite. Nonetheless, for a poor fit which diverges significantly from the true state of the satellite, the errors resulting from the fit become more significant than the intrinsic \ac{TLE} errors, therefore the two metrics should see greater alignment. These expected behaviours are illustrated in \cref{f:error_metric_demonstration}.

\begin{figure}[htb]
    \centering

    \includegraphics{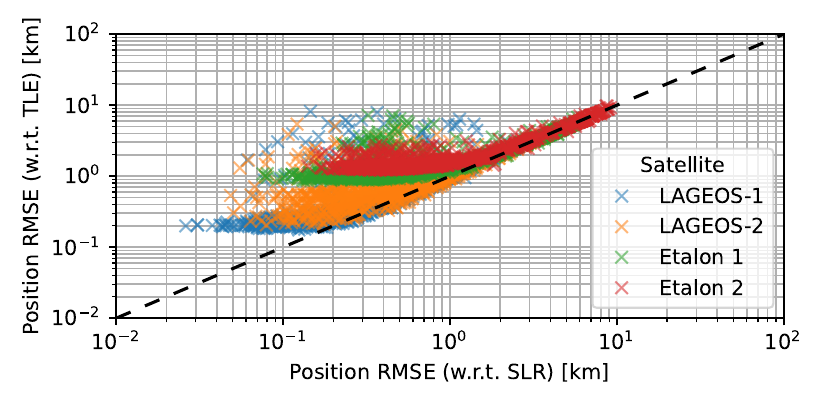}

    \caption{Comparison between \acs{SLR}- and \acs{TLE}-based position \acsp{RMSE} during the 30~days post-fit with linear regressions for each satellite. The dashed line indicates equal \acsp{RMSE}.}
    \label{f:error_comparison}
\end{figure}

A comparison of the \acs{SLR}- and \acs{TLE}-based position \acsp{RMSE} is presented in \cref{f:error_comparison}. The \acs{TLE}-based errors were generally larger than \acs{SLR}-based errors. The \acp{TLE}-based metric had a highly apparent lower bound for each of the four objects, matching the expected behaviour. For larger errors, a more linear behaviour was seen between the metrics.

The results of conducting linear regression on the error metrics, above a minimum error cut-off and in the more linear region, are presented in \cref{t:error_comparison_fits}. Apart from \acs{LAGEOS}-1, the Pearson correlation coefficients ($\rho$) were near unity indicating very strong correlation. Furthermore, the slopes would suggest a near 1:1 relationship between the metrics.

\begin{table}[htb]
    \centering

    \caption{Results of linear regression between the \acs{SLR}- and \acs{TLE}-based metrics for \acs{SLR}-based errors above \SI{1}{\kilo\metre}.}
    \label{t:error_comparison_fits}

    \begin{tabular}{l c c c}
    \toprule
    Satellite   & Slope [-] & Intercept [km]    & $\rho$ [-]    \\
    \midrule
    LAGEOS-1    & 0.81      & 0.53              & 0.43          \\
    LAGEOS-2    & 0.98      & 0.12              & 0.98          \\
    Etalon 1    & 0.91      & 0.40              & 0.98          \\
    Etalon 2    & 0.92      & 0.51              & 0.99          \\
    \bottomrule
    \end{tabular}
\end{table}

Importantly, the very strong correlation in error metrics suggested that the \acs{TLE}-based metric could be used, at the very minimum, for relative comparisons of fit quality for a single object. This result meant that it was feasible to use the \acs{TLE}-based metric for fit quality evaluation of any object in the \ac{TLE} catalogue, allowing testing to be extended to any object meeting the non-manoeuvring requirement.

\subsection{Test Satellite Fits}\label{s:test_fits}

The lack of \enquote{ground truth} data for the test satellites, from either \ac{SLR} or pseudo-ranging, meant that the \acs{TLE}-based error metric had to be used to compare fit quality. This was based on the analysis presented in \cref{s:error_metric_correlation} demonstrating the strong correlation between \acs{SLR}- and \acs{TLE}-based error metrics. Additionally, to smooth out the effect of the variations in \acs{TLE} quality on the reference solution, the prediction duration was extended to 90~days.

\begin{figure}[p]
    \centering
    \includegraphics{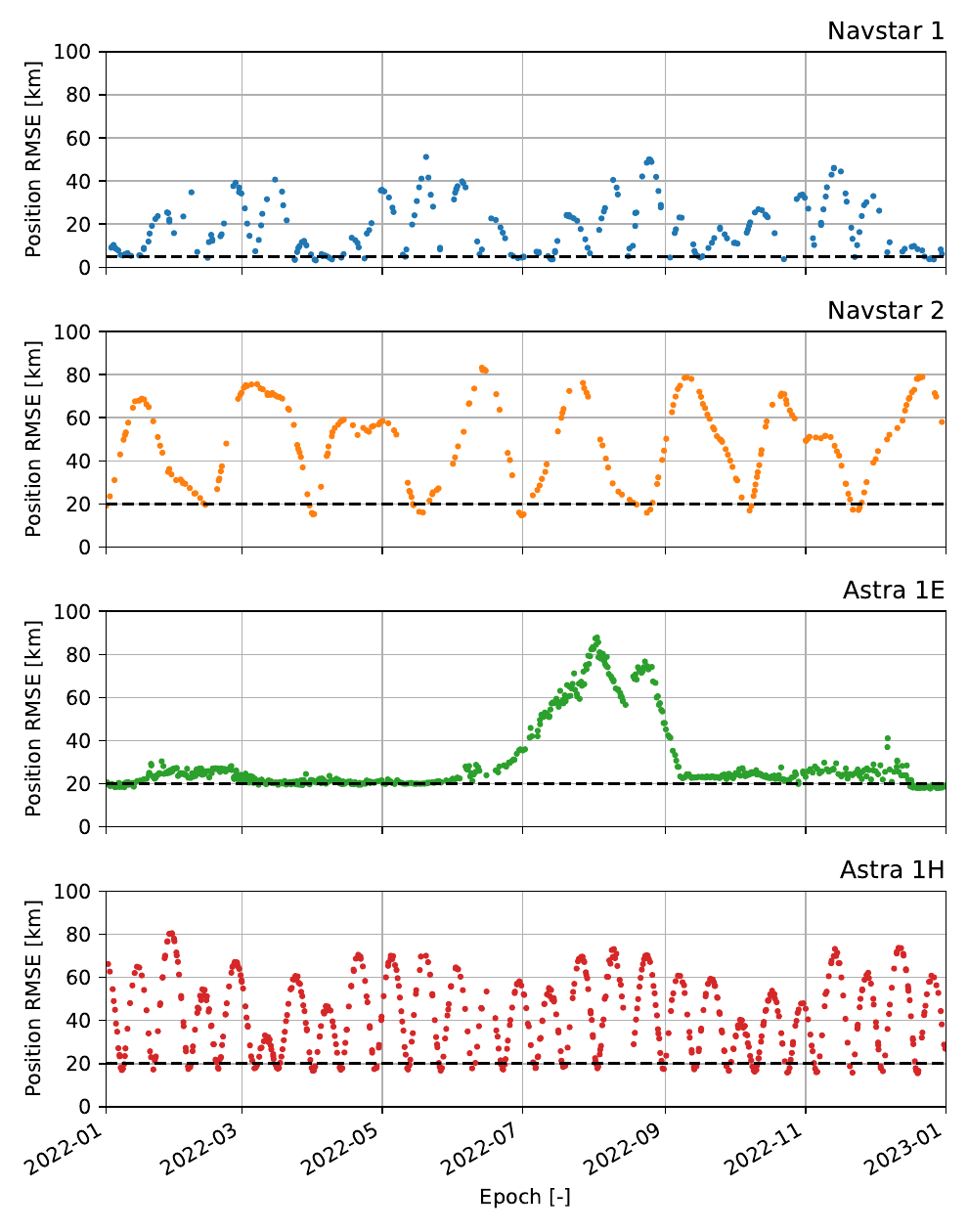}
    \caption{Position \acs{RMSE}, with respect to \acsp{TLE}, during 90~days post-epoch for test satellite \acsp{TLE}. The dashed lines represent the established baselines for each satellite.}
    \label{f:tle_rmse_test}
\end{figure}

The quality of the test satellites' \acp{TLE}, relative to future \acp{TLE}, was evaluated to establish a benchmark for each satellite, as in \cref{s:calibration_fits}. The results are presented in \cref{f:tle_rmse_test}. Navstar~1 and~2, and Astra~1H showed periodic behaviour in error throughout the year, although seemingly each with different structures and periods. In the case of Navstar~1, the errors remained below \SI{50}{\kilo\metre} throughout the year. In the case of Navstar~2 and Astra~1H, the errors oscillated in a band from \SIrange{20}{80}{\kilo\metre}. Astra~1E had errors which remained around \SI{20}{\kilo\metre} consistently with little variation, apart from a period between July and September where the error grew significantly to \SI{90}{\kilo\metre}.

\begin{figure}[p]
    \centering

    \begin{subfigure}[c]{0.5\textwidth}
        \centering
        \includegraphics{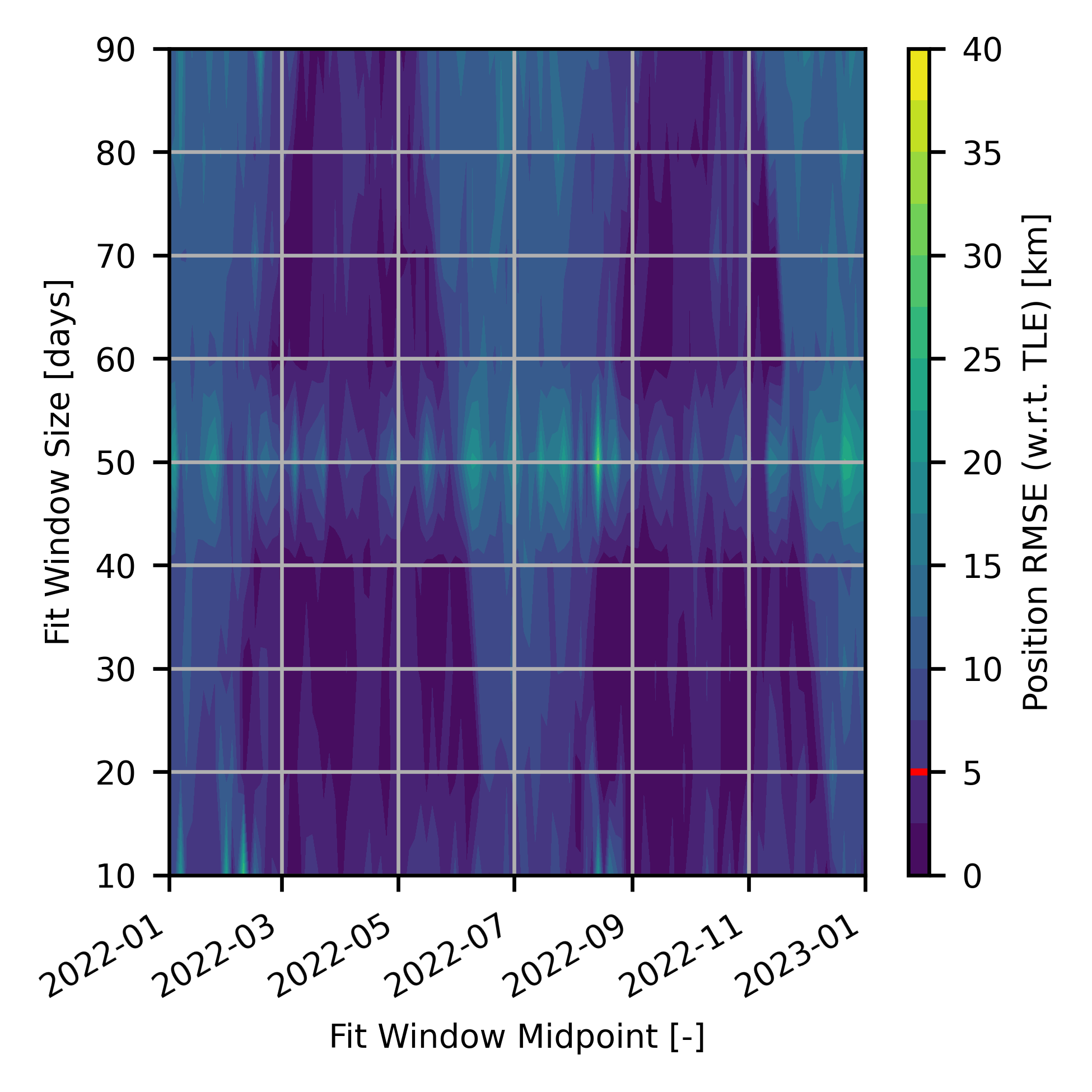}
        \caption{Navstar~1.}
    \end{subfigure}%
    \begin{subfigure}[c]{0.5\textwidth}
        \centering
        \includegraphics{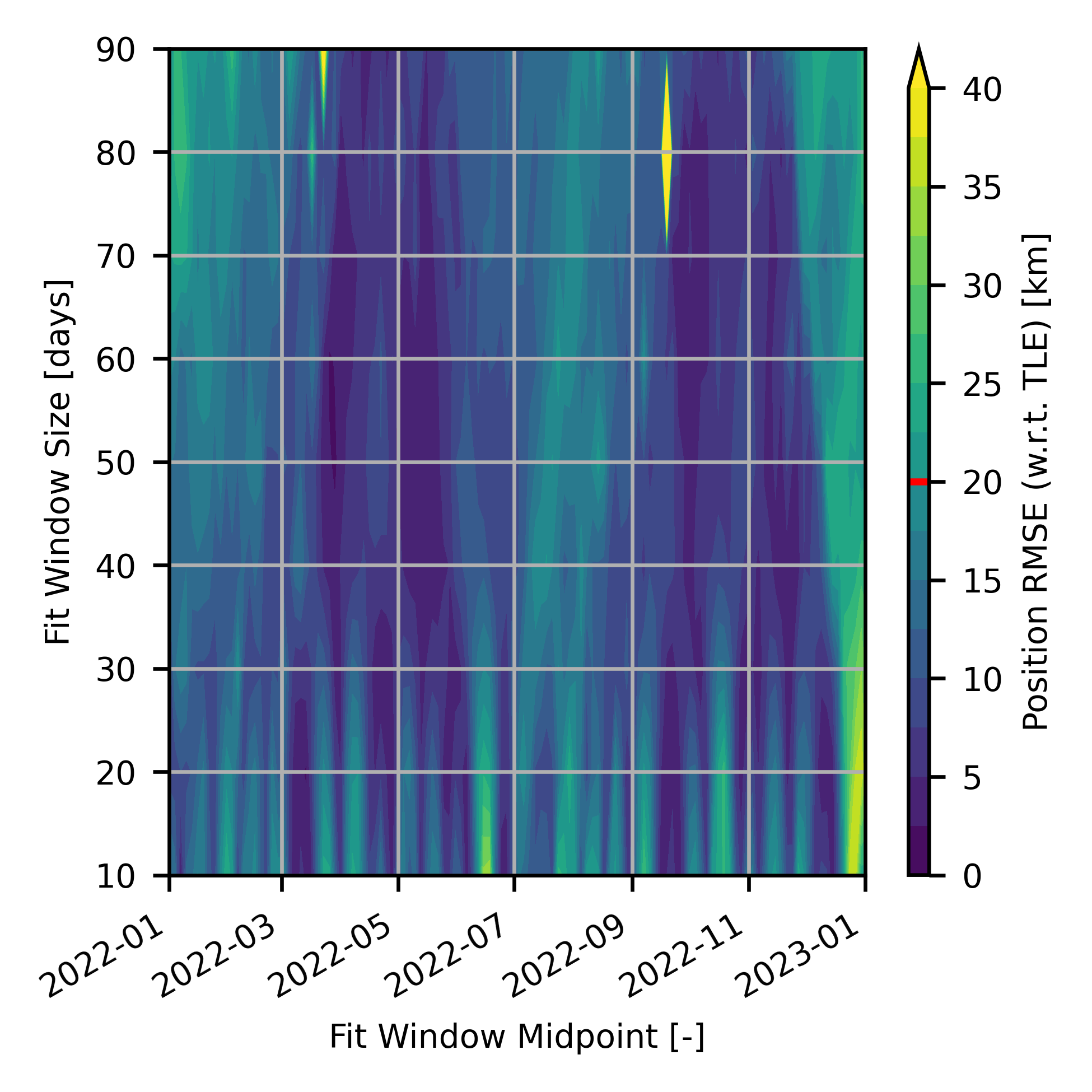}
        \caption{Navstar~2.}
    \end{subfigure}

    \vspace{0.5\baselineskip}

    \begin{subfigure}[c]{0.5\textwidth}
        \centering
        \includegraphics{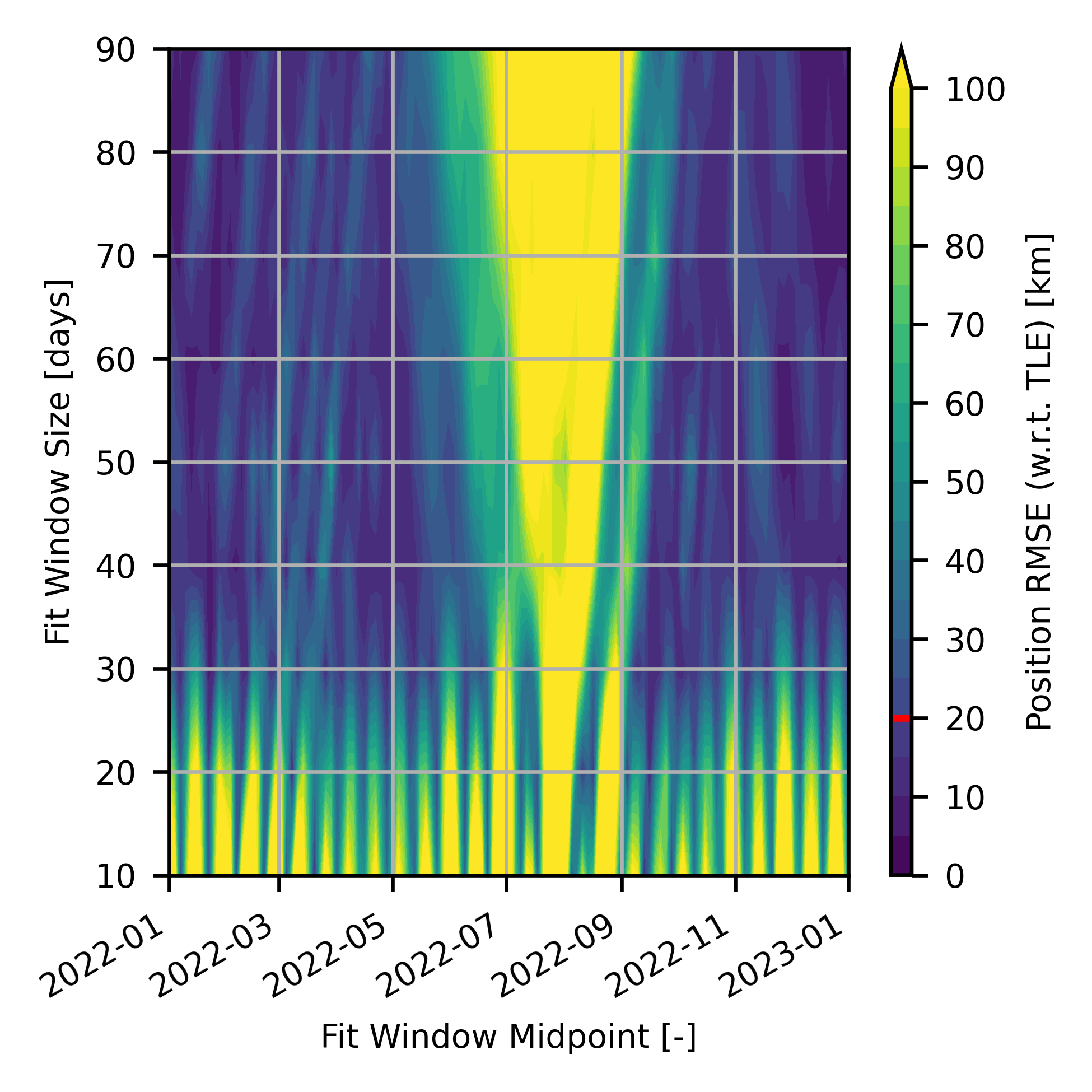}
        \caption{Astra~1E.}
    \end{subfigure}%
    \begin{subfigure}[c]{0.5\textwidth}
        \centering
        \includegraphics{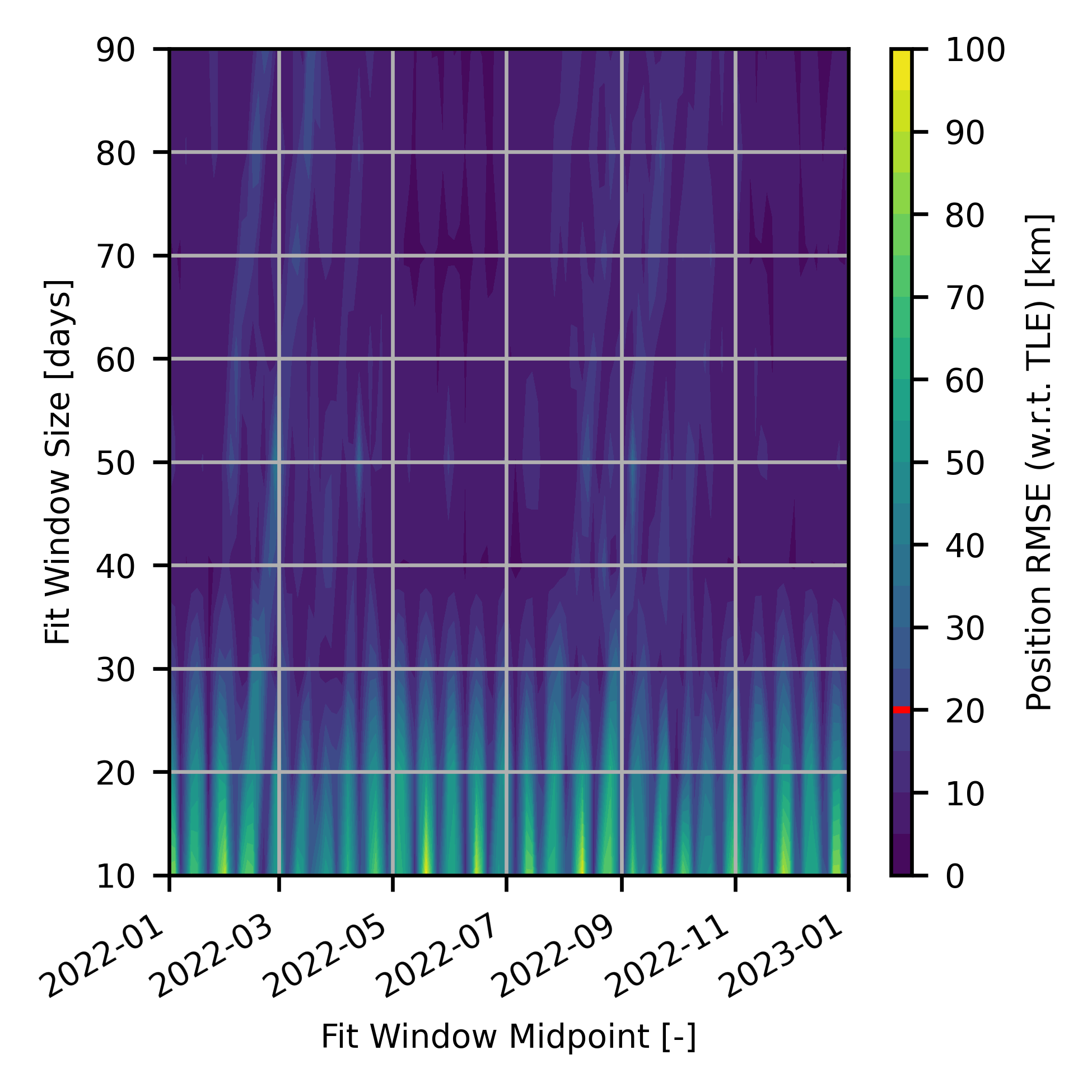}
        \caption{Astra~1H.}
    \end{subfigure}

    \vspace{0.25\baselineskip}
    
    \caption{Position \acs{RMSE} during the 90~days post-fit for long-term biased fits. The red marker in the colour bar indicates the \acs{RMSE} baseline derived from \acsp{TLE}.}
    \label{f:test_long_biased}
\end{figure}

The results of fits on the test satellites are presented in \cref{f:test_long_biased}. In general, fits on the four objects showed large domains where the fit performance meets or exceeds the benchmark, suggesting that the fits were successfully reducing the prediction errors with respect to propagating the \acp{TLE} directly. Increasing the fit window size was generally beneficial, although the error behaviour was more complex than for test objects and fits were highly dependent on the quality of the \acp{TLE} within and following the fit window.

The periodic variation in error throughout the year for short fit windows was highly apparent for Navstar~2, and Astra~1E and~1H. The fits on Navstar~1, on the other hand, did not show this variation as strongly, instead demonstrating a longer period variation throughout the year with slightly increased errors at two points in the year, November through February and June through August. Nevertheless, along with Astra~1H, a notable increase in periodic error was visible for 50-day windows, a duration close to twice the Moon's orbital period. The increased \ac{TLE} error for Astra~1E between July and September was highly apparent, as seen by very high errors for the fits during the time period. It was unexpected that Navstar~1 would not show oscillatory behaviour in the fits despite oscillations in \ac{TLE} quality, and vice versa for Astra~1E, which showed oscillatory behaviour for short-term windows despite constant \acs{TLE}-on-\acs{TLE} error.

\begin{figure}[htbp]
    \centering
    \includegraphics{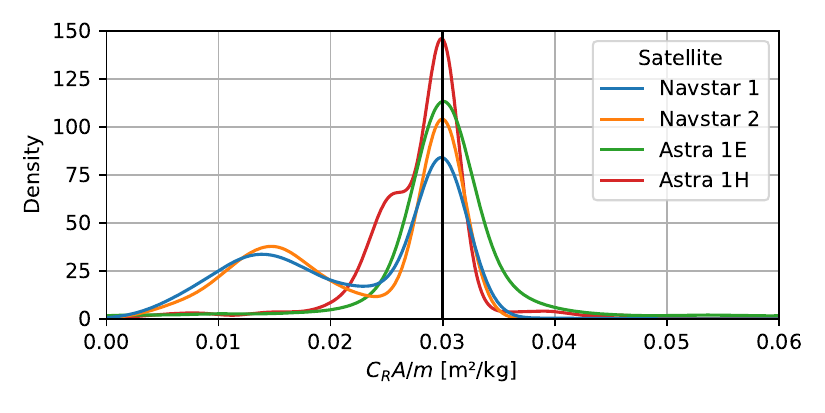}
    \caption{\acs{SRP} coefficient distributions for the test satellites fits.}
    \label{f:srp_coefficients}
\end{figure}

The distributions of the estimated \ac{SRP} coefficients are presented in \cref{f:srp_coefficients}. For all four objects, the majority of fits resulted in coefficients close to the initial guess of \SI{3e-2}{\metre^2/\kilogram}. It was expected that the majority of estimated \ac{SRP} coefficients would remain in the region of the initial guess due to the aforementioned low observability of the parameter during the \ac{P-OD} process. In the case of Navstar~1 and~2, a second peak was visible around \SI{1.5e-2}{\metre^2/\kilogram}.
 
\section{Debiased P-OD}

\subsection{Calibration Satellite Fits}\label{s:cal_debiased_fits}

\begin{figure}[p]
    \centering

    \begin{subfigure}[c]{0.5\textwidth}
        \centering
        \includegraphics{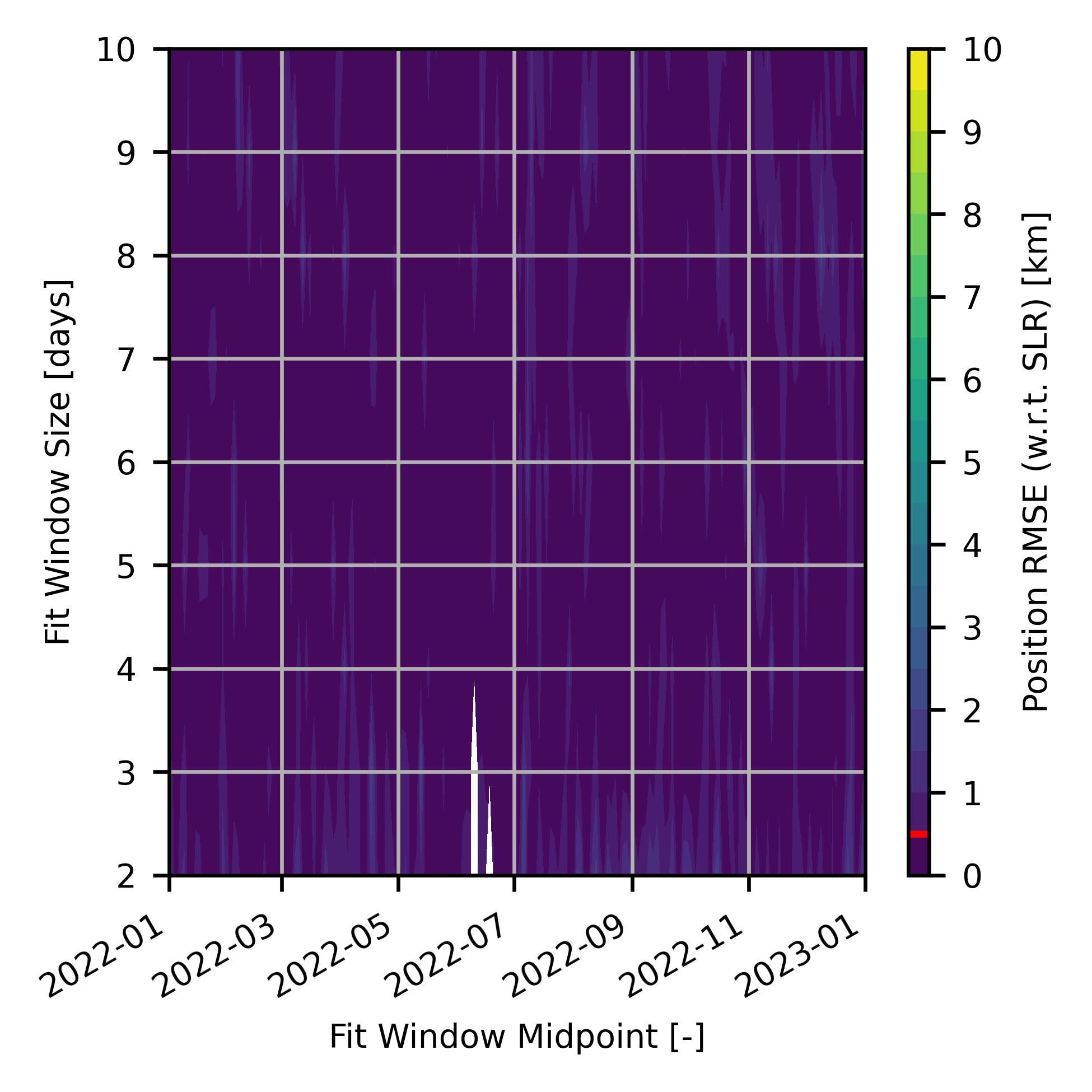}
        \caption{\acs{LAGEOS}-1.}
    \end{subfigure}%
    \begin{subfigure}[c]{0.5\textwidth}
        \centering
        \includegraphics{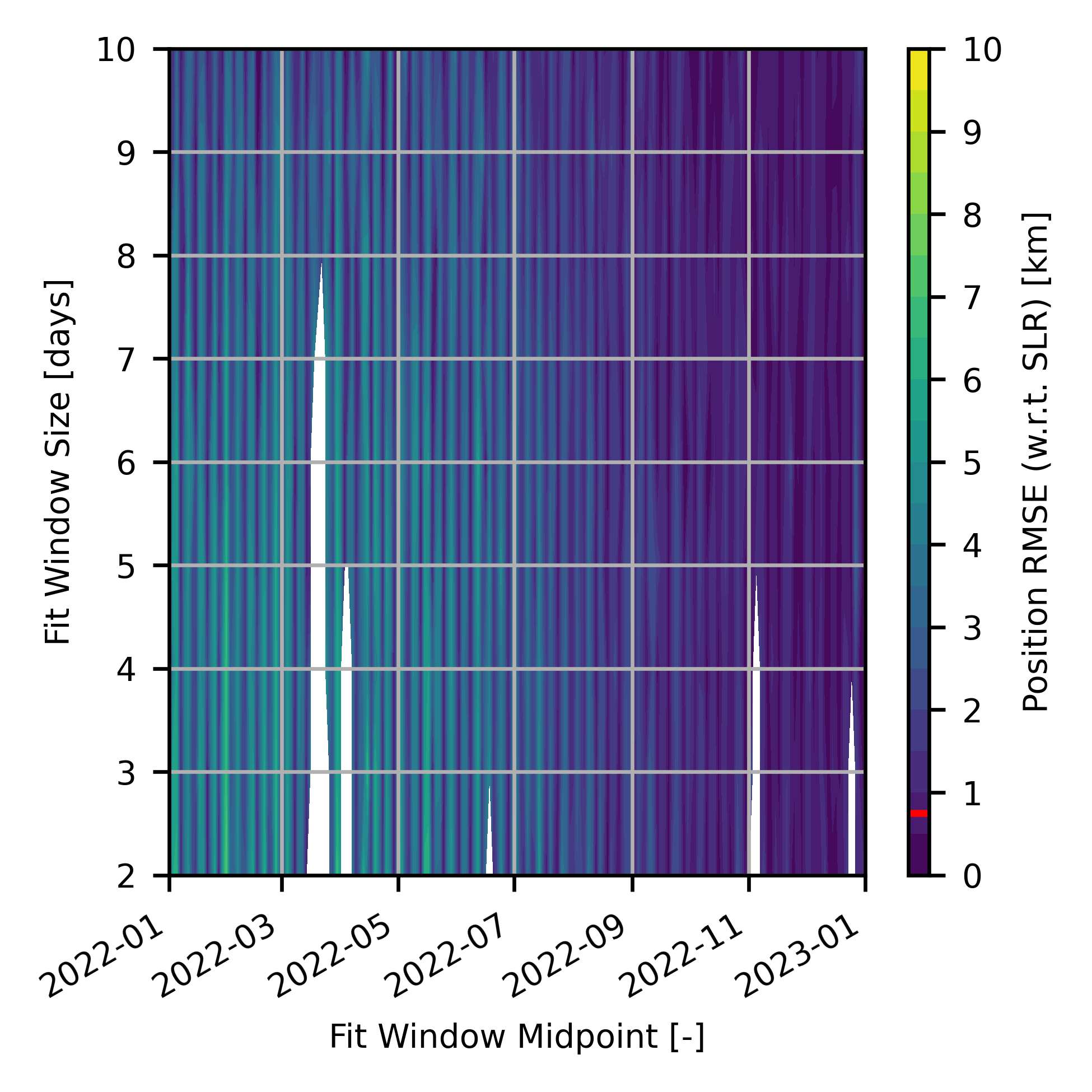}
        \caption{\acs{LAGEOS}-2.}
    \end{subfigure}

    \vspace{0.5\baselineskip}

    \begin{subfigure}[c]{0.5\textwidth}
        \centering
        \includegraphics{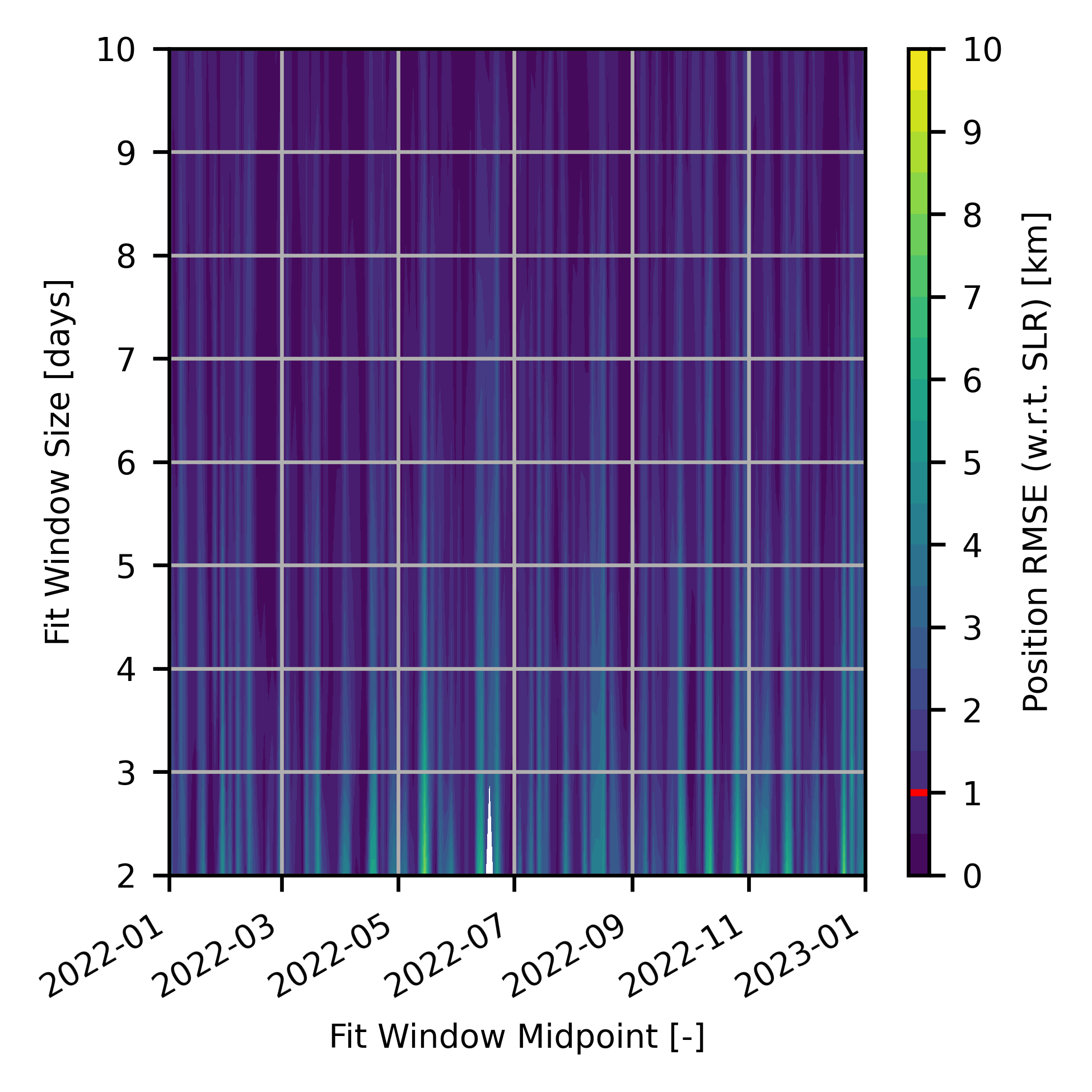}
        \caption{Etalon~1.}
    \end{subfigure}%
    \begin{subfigure}[c]{0.5\textwidth}
        \centering
        \includegraphics{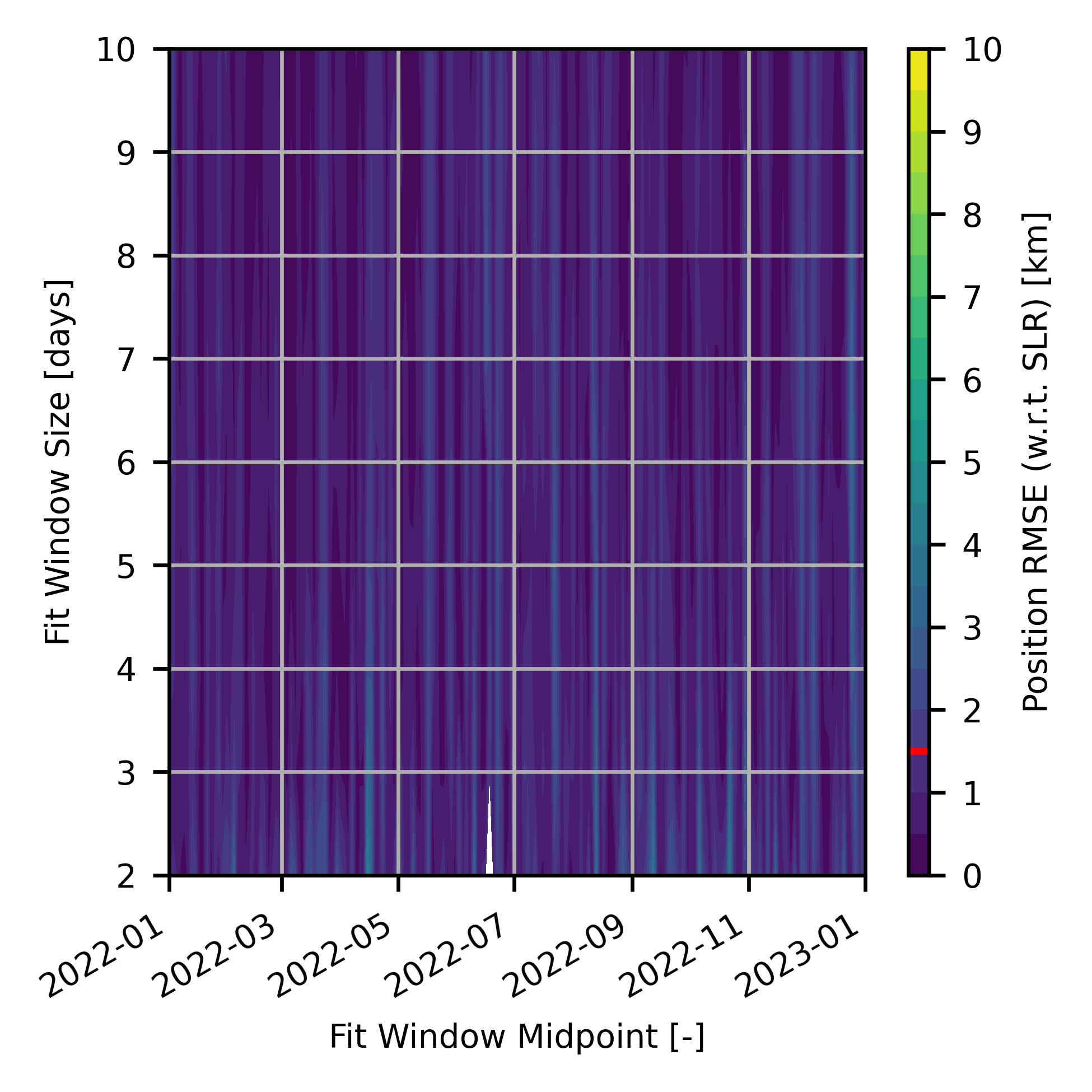}
        \caption{Etalon~2.}
    \end{subfigure}

    \vspace{0.25\baselineskip}
    
    \caption{Position \acs{RMSE} during the 30~days post-fit for short-term debiased fits. The red marker in the colour bar indicates the \acs{RMSE} baseline derived from \acsp{TLE}. Note that white regions indicate failed cases due to an insufficient number of \acsp{TLE}.}
    \label{f:cal_short_debiased}
\end{figure}

\begin{figure}[p]
    \centering

    \begin{subfigure}[c]{0.5\textwidth}
        \centering
        \includegraphics{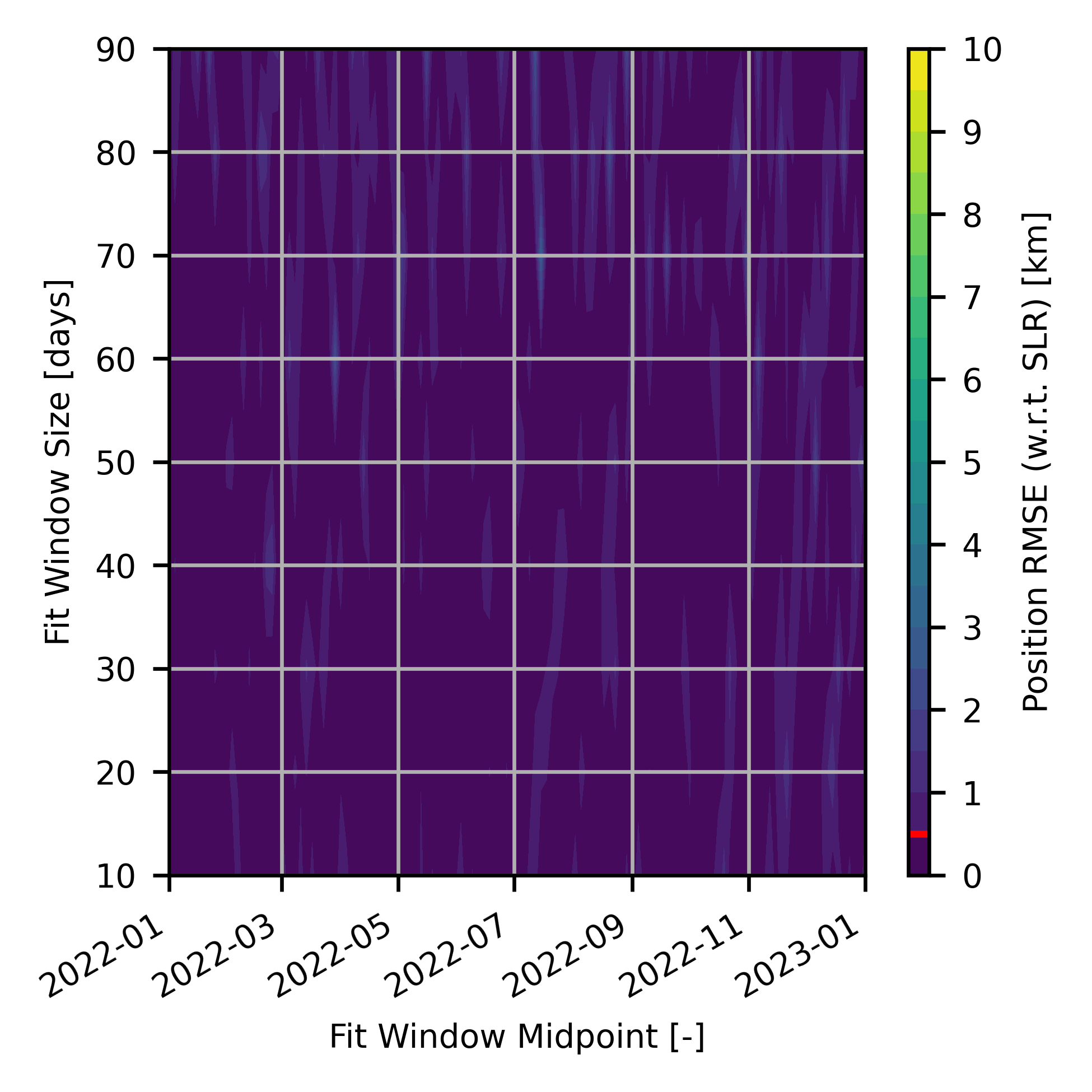}
        \caption{\acs{LAGEOS}-1.}
    \end{subfigure}%
    \begin{subfigure}[c]{0.5\textwidth}
        \centering
        \includegraphics{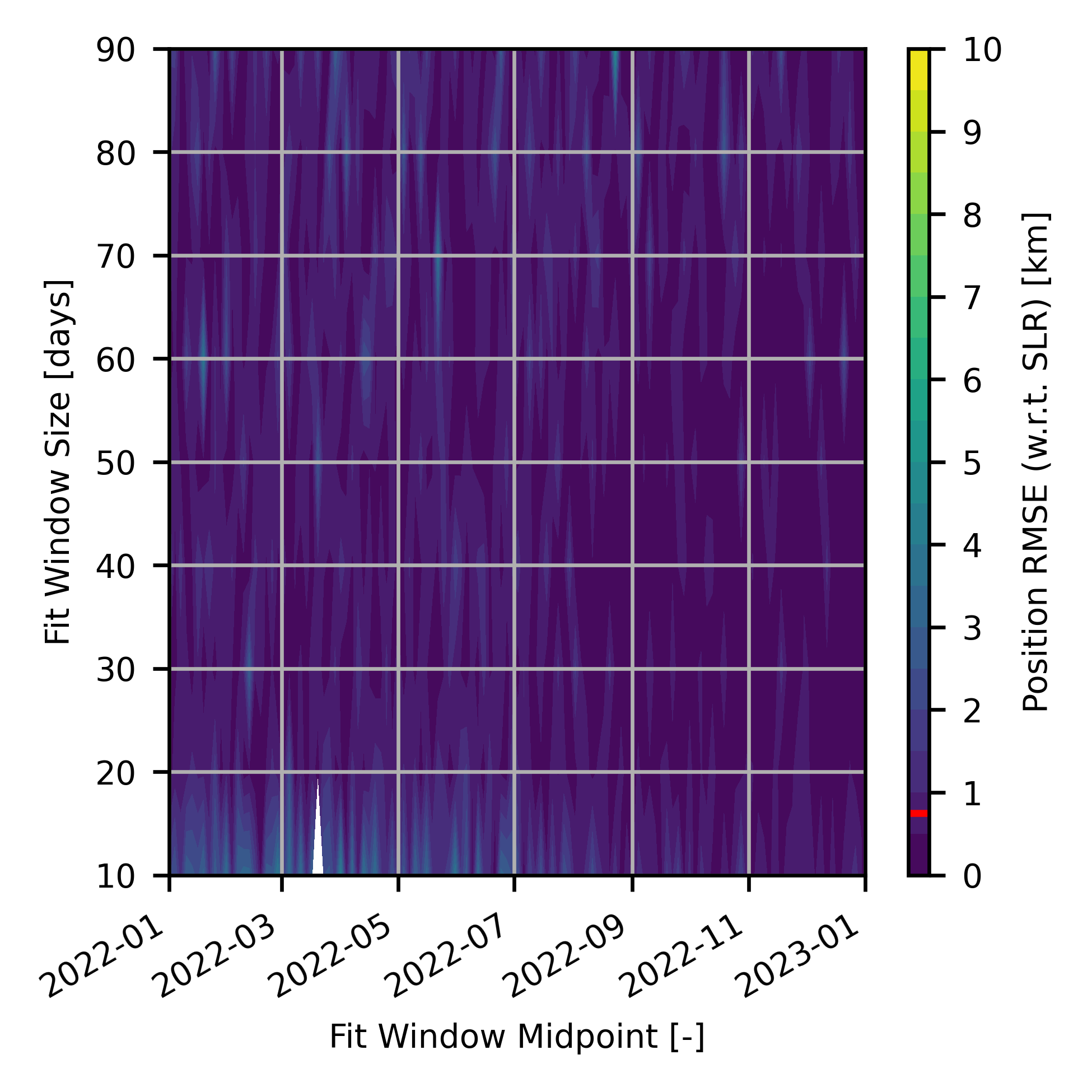}
        \caption{\acs{LAGEOS}-2.}
    \end{subfigure}

    \vspace{0.5\baselineskip}

    \begin{subfigure}[c]{0.5\textwidth}
        \centering
        \includegraphics{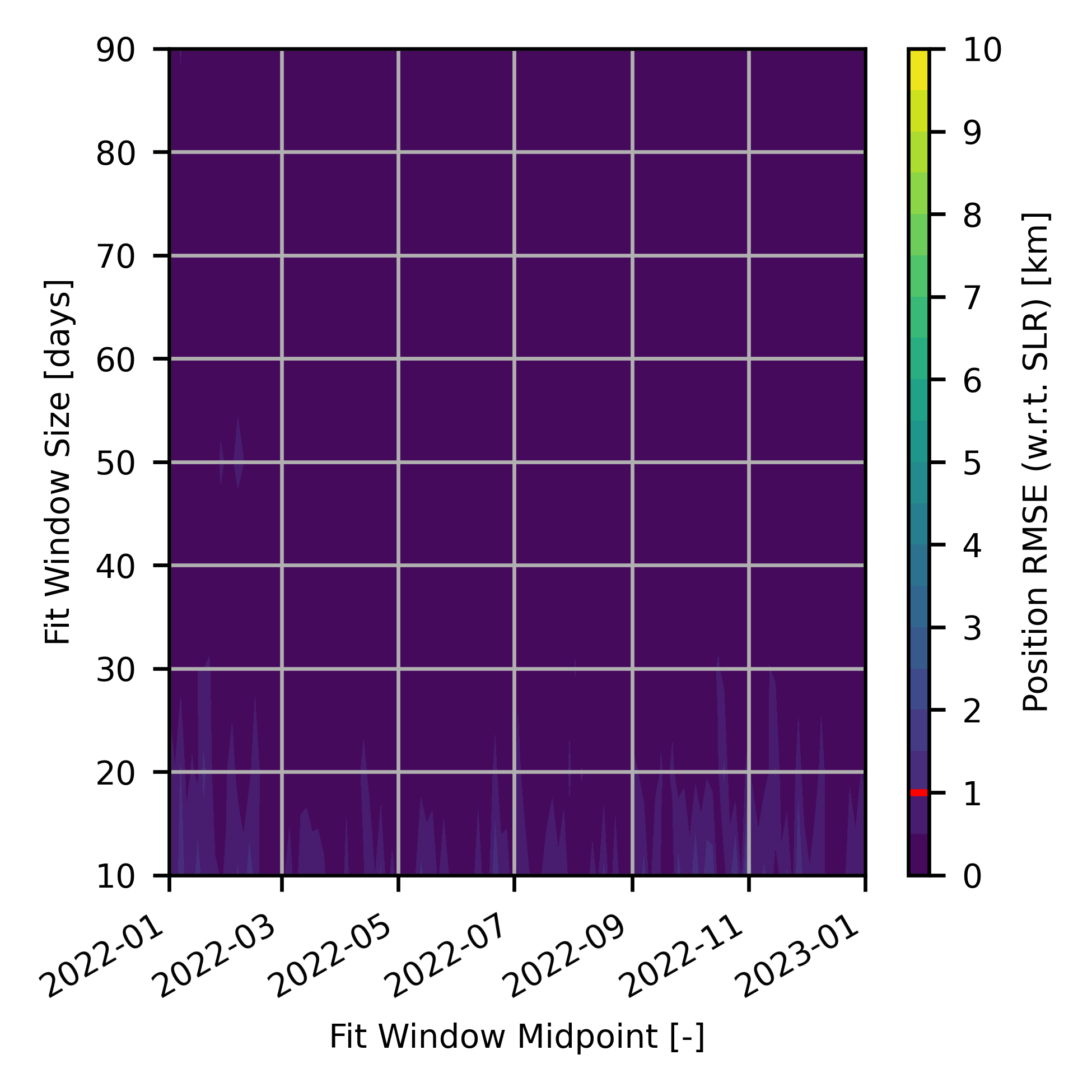}
        \caption{Etalon~1.}
    \end{subfigure}%
    \begin{subfigure}[c]{0.5\textwidth}
        \centering
        \includegraphics{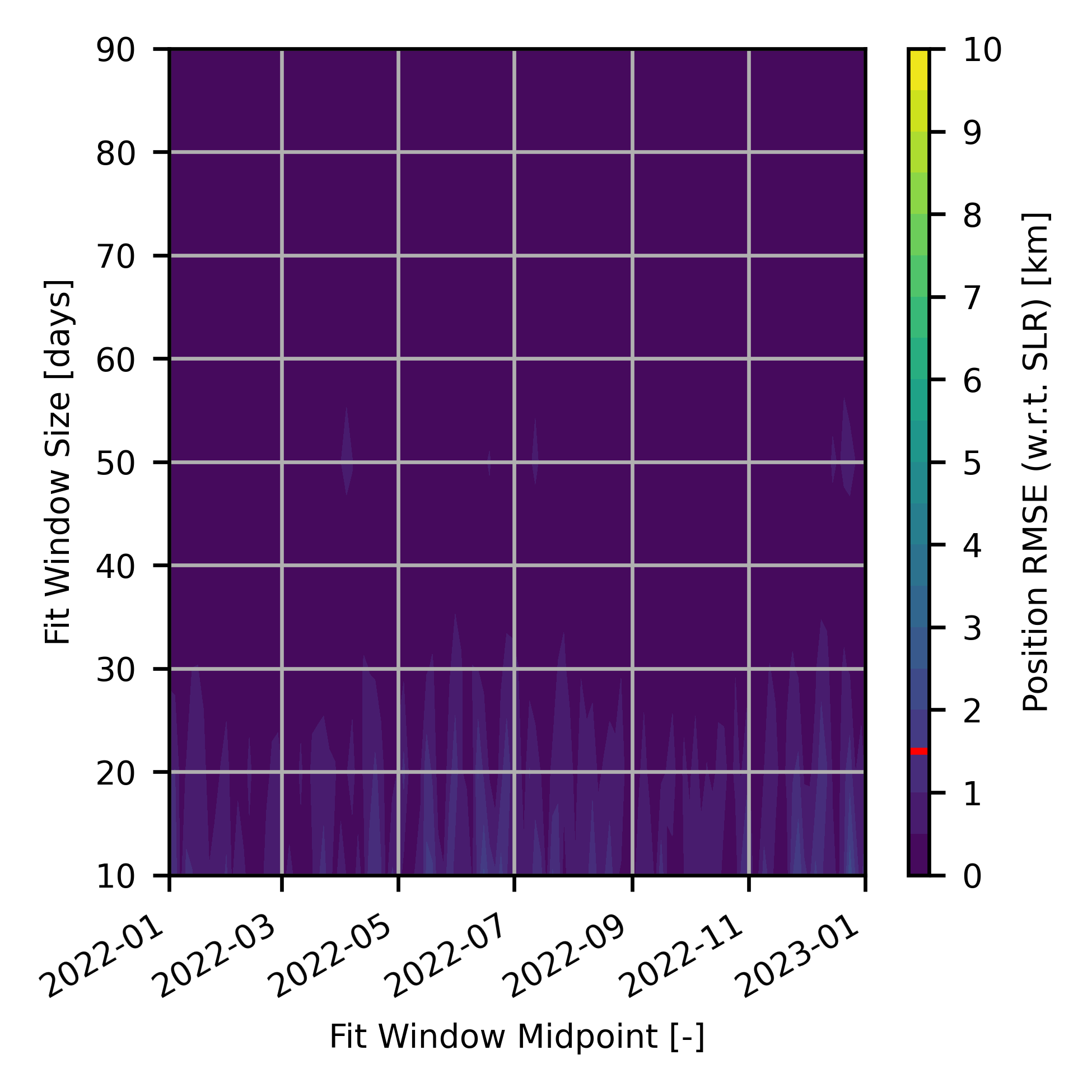}
        \caption{Etalon~2.}
    \end{subfigure}

    \vspace{0.25\baselineskip}
    
    \caption{Position \acs{RMSE} during the 30~days post-fit for long-term debiased fits. The red marker in the colour bar indicates the \acs{RMSE} baseline derived from \acsp{TLE}. Note that white regions indicate failed cases due to an insufficient number of \acsp{TLE}.}
    \label{f:cal_long_debiased}
\end{figure}

The results of fits on the test satellites, following debiasing as described in \cref{s:debiasingmethod}, are presented in \cref{f:cal_short_debiased,f:cal_long_debiased}. The fits on \acs{LAGEOS}-1 saw minor reductions in error with some suppression of the periodic behaviour for short fit windows. The results for \acs{LAGEOS}-2 remained mainly unaffected, demonstrating the same error behaviour as the fits on the biased pseudo-observations. This was expected due to the corresponding debiasing model have very little impact on the along-track position. The fits on Etalon~1 and~2 saw the most significant changes in error: although the periodic variations were still visible, the peak errors were reduced from approximately \SI{10}{\kilo\metre} to \SI{2}{\kilo\metre}.

\begin{figure}[htb]
    \centering

    \begin{subfigure}[c]{0.5\textwidth}
        \centering
        \includegraphics{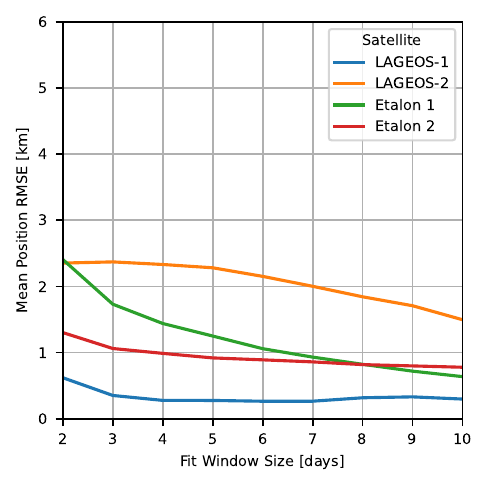}
        \caption{Short-term.}
    \end{subfigure}%
    \begin{subfigure}[c]{0.5\textwidth}
        \centering
        \includegraphics{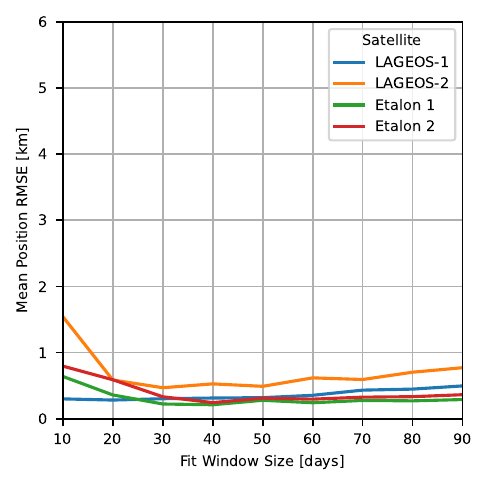}
        \caption{Long-term.}
    \end{subfigure}

    \vspace{0.25\baselineskip}
    
    \caption{Mean position \acs{RMSE} during the 30~days post-fit for debiased fits.}
    \label{f:cal_mean_debiased}
\end{figure}

The suppression of the oscillations in error was seen in the reduction of the mean position \ac{RMSE} in \cref{f:cal_mean_debiased}. For short-term fits of Etalon~1, the mean position error was reduced from an approximately constant \SI{3.5}{\kilo\metre} to a trend decreased from approximately \SIrange{2.5}{0.5}{\kilo\metre}, representing an approximate \SIrange{30}{85}{\%} reduction in error. For Etalon~2, the mean error was reduced from approximately \SI{5}{\kilo\metre} to \SI{1}{\kilo\metre}, an \SI{80}{\%} reduction. For the long-term fits, in addition to the mean position errors converging sooner for a small fit window size, the peaks at 50 and 80~days were removed with mean errors remaining below \SI{1}{\kilo\metre} after fit windows larger than 20~days.

\subsection{Effect of Fit Window on Bias Impact}

The influence of window size on the bias can be analysed by considering the mean bias during the window. For the simplified sinusoidal model, \cref{e:bias_model}, this can be found through direct, analytical integration:
\begin{align}
    \overline{\Delta \theta_T} (t^*, \Delta t) &= \frac{1}{\Delta t} \int_{t^* - \frac{\Delta t}{2}}^{t^* + \frac{\Delta t}{2}} \Delta \theta_T (t) \ dt \text{,} \\[0.75em]
    &\approx \frac{1}{\Delta t} \int_{t^* - \frac{\Delta t}{2}}^{t^* + \frac{\Delta t}{2}} a \sin\left( \frac{2 \pi}{b} \left(t + c\right) \right) + d \ dt \text{,} \\[0.75em]
    &\approx \frac{ab}{\pi \Delta t} \sin\left( \frac{2 \pi}{b} \left( t^* + c \right) \right) \sin \left( \frac{\pi \Delta t}{b} \right) + d \text{,} \label{e:mean_transverse_error}
\end{align}
where $\overline{\Delta \theta_T}$ is the mean along-track angular error, $t^*$ is the mid-point of the window, and $\Delta t$ its size.

The same method can be used to find the mean bias error rate during the window:
\begin{align}
    \overline{\Delta \dot{\theta}_T} (t^*, \Delta t) &= \frac{1}{\Delta t} \int_{t^* - \frac{\Delta t}{2}}^{t^* + \frac{\Delta t}{2}} \Delta \dot{\theta}_T (t) \ dt \text{,} \\[0.75em]
    &= \frac{1}{\Delta t} \left[ \vphantom{\frac{1}{1}} \Delta \theta_T (t) \right]_{t^* - \frac{\Delta t}{2}}^{t^* + \frac{\Delta t}{2}} \text{,} \\[0.75em]
    &\approx \frac{1}{\Delta t} \left[ a \sin\left( \frac{2 \pi}{b} \left(t + c\right) \right) + d \right]_{t^* - \frac{\Delta t}{2}}^{t^* + \frac{\Delta t}{2}} \text{,} \\[0.75em]
    &\approx \frac{2 a}{\Delta t} \cos \left( \frac{2 \pi}{b} \left(t^* + c\right) \right) \sin \left( \frac{\pi \Delta t}{b} \right) \text{,} \label{e:mean_transverse_error_rate}
\end{align}
where $\overline{\Delta \dot{\theta}_T}$ is the mean along-track angular error rate.

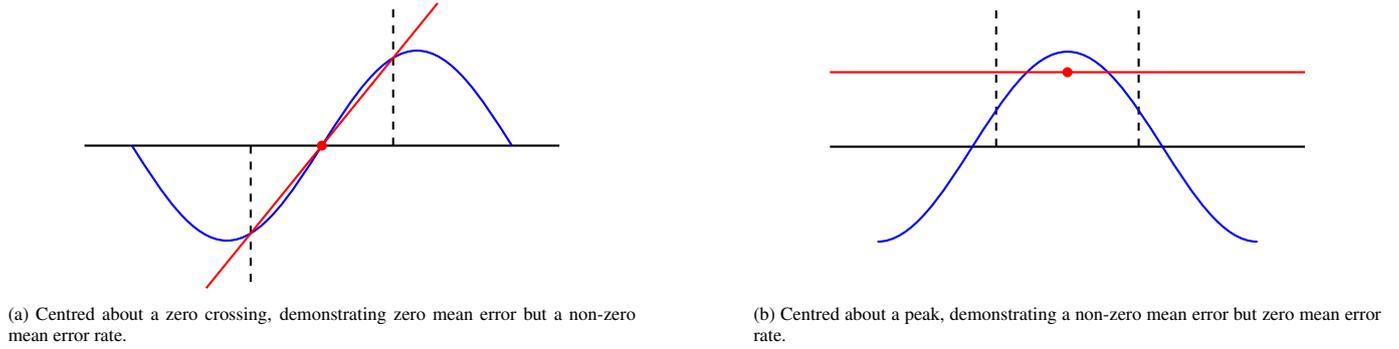
\begin{figure}[htb]
    \centering

    \vspace{0.5\baselineskip}

    \begin{subfigure}[t]{0.45\textwidth}
        \centering
        \begin{tikzpicture}[scale=1.25]
            \draw[black, thick] (-2.5,0) -- (2.5, 0);

            \draw[black, thick, dashed] (-0.75, 0) -- (-0.75, -1.5);
            \draw[black, thick, dashed] (0.75, 0) -- (0.75, 1.5);
        
            \draw[blue, thick] (-2,0) sin (-1,-1) cos (0,0) sin (1,1) cos (2,0);

            \draw[red, thick] (-1.21769, -1.5) -- (1.21769, 1.5);

            \fill[red] (0, 0) circle[radius=1.5pt];

            \path (-2.5, -1.5) -- (2.5, 1.5);
        \end{tikzpicture}

        \caption{Centred about a zero crossing, demonstrating zero mean error but a non-zero mean error rate.}
    \end{subfigure}
    \hspace{0.075\textwidth}
    \begin{subfigure}[t]{0.45\textwidth}
        \centering
        \begin{tikzpicture}[scale=1.25]
            \draw[black, thick] (-1.5,0) -- (3.5, 0);

            \draw[black, thick, dashed] (0.25, 0) -- (0.25, 1.5);
            \draw[black, thick, dashed] (1.75, 0) -- (1.75, 1.5);
        
            \draw[blue, thick] (-1,-1) cos (0,0) sin (1,1) cos (2,0) sin (3,-1);

            \draw[red, thick] (-1.5, 0.78421) -- (3.5, 0.78421);

            \fill[red] (1, 0.78421) circle[radius=1.5pt];

            \path (-1.5, -1.5) -- (3.5, 1.5);
        \end{tikzpicture}

        \caption{Centred about a peak, demonstrating a non-zero mean error but zero mean error rate.}
    \end{subfigure}

    \vspace{0.25\baselineskip}
    
    \caption{Example of averaging a bias model (blue), illustrated by considering a line of best fit (red) for a restricted domain (black, dashed).}
    \label{f:bias_demonstration}
\end{figure}

A key feature of the mean error is that the period and phase remain unchanged by the averaging process, remaining independent of the fit window size. This would explain why the variations in position \ac{RMSE} showed the same pattern as the fit window size was increased. The mean error rate, on the other hand, is phase-shifted by \SI{90}{deg} with respect to the mean error. Putting aside the influence of the sine term for one moment, this would suggest that the amplitude of the mean error rate is highest when the mean error is minimised and vice-versa, as illustrated in \cref{f:bias_demonstration}.

Increasing the size of the fit window has important effects on the mean error and mean error rate. The amplitude of the periodic terms are reduced due to the inclusion of the reciprocal of the fit window size as a coefficient. In the limit, this results in the periodic behaviour being removed entirely:
\begin{align}
    \lim_{\Delta t \to \infty} \overline{\Delta \theta_T} (t^*, \Delta t) &\approx d \text{,} \\[0.5em]
    \lim_{\Delta t \to \infty} \overline{\Delta \dot{\theta}_T} (t^*, \Delta t) &\approx 0 \text{.}
\end{align}

Alternatively, the mean error and error rate can be minimised by considering the sinusoidal component present in both \cref{e:mean_transverse_error,e:mean_transverse_error_rate}:
\begin{align}
    \sin \left( \frac{\pi \Delta t}{b} \right) &= 0 \text{,} \\[0.75em]
    \frac{\pi \Delta t}{b} &= \pi n \text{,} \\[0.75em]
    \frac{\Delta t}{b} &= n \text{,}
\end{align}
where $n$ is an integer. In other words, the mean along-track angular error and error rate are minimised when the fit window duration is an integer multiple of the bias period. It is worth noting that this does not match the behaviour seen in the test satellite fits, where the position \ac{RMSE} appeared to increase rather than decrease near the integer multiples. A potential cause of this discrepancy might include the additional harmonics seen in the \ac{TLE} data, or the need for a more accurate bias model.

The effect of extending the fit window is to convert short-term biases into long-term noise. By capturing multiple cycles of the bias, with a period of approximately 27.5~days as identified in \cref{s:debiasingmethod}, the mean along-track error averages to near-zero. This is due to times of the \acp{TLE} leading and lagging the true position cancelling out. This results in a solution to the batch least squares process which is not strongly influenced by the bias in a small part of the cycle.

\subsection{Alternative Bias Models}

The main limitation of the along-track error model was that it was not generalised for any satellite, a shortcoming present in many of the models proposed in previous literature. Fundamentally, this is a result of creating models based on a small subset of objects with known precise ephemerides which may not represent the wider catalogue.

In the case of the model presented above, amplitudes varied by satellite, and \acs{LAGEOS}-1 showed slightly different behaviour to the Etalon satellites. The outlying behaviour of \acs{LAGEOS}-2 raises the question of whether other satellites will have similar behaviour, and whether it is a direct result of the SGP4 model, the dynamics of specific orbital regimes, or additional non-physical factors introduced during the \ac{TLE} generation process. Additional components of the state vector may benefit also from debiasing: although not implemented here, the radial velocity was found to have similar periodic variations with respect to epoch.

The key for future bias model development lies with the careful selection of which parameters are included as inputs. These can dramatically affect both the model's performance and whether it is generalisable to a larger proportion of the \ac{TLE} catalogue. For example, the lunar perturbation is a product of the distance between the satellite and the Moon, therefore it may be appropriate to include the Moon's position as a parameter in a bias model. In the case of \cref{e:bias_model}, the use of time could be seen as a proxy for the Moon's position, supported by the fitted models' period closely matching that of the Moon.

An alternative bias model, based on the Moon's position along its orbit is presented:
\begin{equation}
    \Delta \theta_T(M) \approx a \sin (M + b) \text{,}
\end{equation}
where $M$ is the Moon's mean anomaly, and $a$ and $b$ are the bias amplitude and phase offset respectively.

While this model adds some complexity over \cref{e:bias_model}, due to requiring ephemerides for the Moon, it is still relatively simple computationally. Furthermore, it is linked more directly to the suspected source of the bias, mismodelling of the lunar perturbation, than indirectly via time.

The fitted parameters for the anomaly-based bias model are presented in \cref{t:moon_model_parameters}. As expected, the amplitudes of the model are very similar to those presented in \cref{t:transverse_errors_model_parameters}, including the very small value for \acs{LAGEOS}-2, suggesting a similarly poor fit. Notably, the phase offsets for \acs{LAGEOS}-1, and Etalon 1 and 2 are very similar, lying within a \SI{0.26}{rad} band.

\begin{table}[htb]
    \centering

    \caption{Fitted parameters for the anomaly-based bias model.}
    \label{t:moon_model_parameters}

    \sisetup{exponent-mode=input}

    \begin{tabular}{l S[table-format=1.2e+1] S[table-format=+1.2e+1]}
    \toprule
    Satellite       & {$a$ [rad]}   & {$b$ [rad]}   \\
    \midrule
    LAGEOS-1        & 1.33e-5       & 1.73e-1       \\
    LAGEOS-2        & 1.82e-6       & -3.23e-1      \\
    Etalon 1        & 4.62e-5       & -8.40e-2      \\
    Etalon 2        & 6.59e-5       & 9.97e-3       \\
    \bottomrule
    \end{tabular}
\end{table}

\begin{figure}[p]
    \centering
    \includegraphics{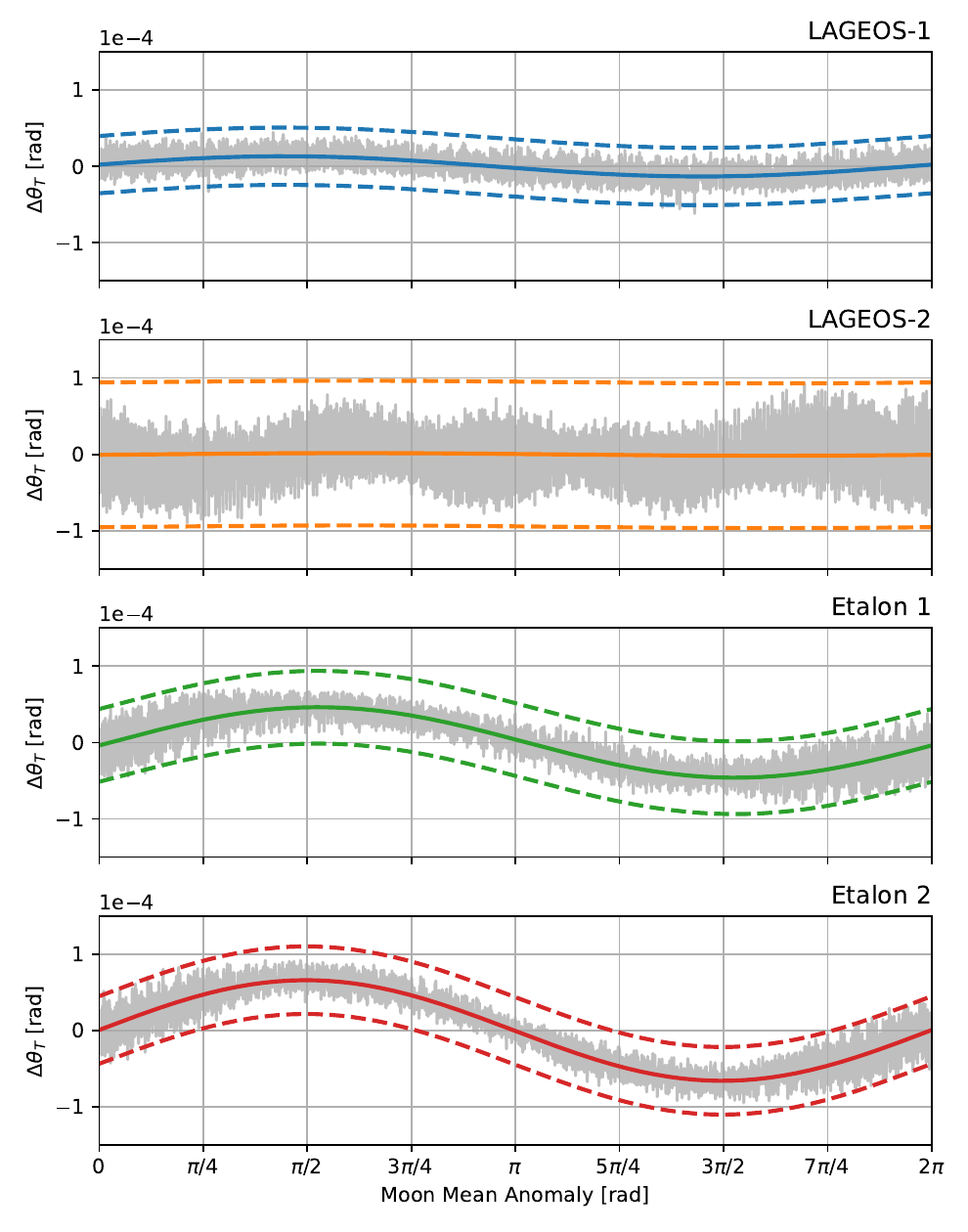}
    \caption{Along-track bias model, based on the Moon's mean anomaly. The \acs{SLR} data and fitted models are presented in grey and colour respectively, with the dashed lines representing the $3\sigma$ bounds.}
    \label{f:moon_model}
\end{figure}

The anomaly-based bias models are illustrated in \cref{f:moon_model}. The model for \acs{LAGEOS}-2 can be seen to match poorly with the variations in the along-track bias, appearing similar to a horizontal line through zero. The variations in along-track bias for \acs{LAGEOS}-2 appear to have higher frequency components, suggesting that higher frequency terms should be considered. For \acs{LAGEOS}-1, and Etalon 1 and 2 the models show good agreement visually with all variations lying within the $3\sigma$ bounds. In the case of Etalon 1 and 2, the amplitude of the noise about the model appears to increase in the region near perilune. This would be logical given the increased magnitude of the perturbation as the Moon makes its closest approach to Earth and the satellites.

Data-driven solutions, such as the simplified models presented in this study or other machine learning methods~\citep{ly_correcting_2020}, are limited to the availability of \enquote{ground truth} data, such as from \ac{SLR} or post-processed pseudoranging, for training, validation, and testing. Furthermore, for objects with high quality \enquote{ground truth} data, it is unknown whether these higher accuracy products are used during the \ac{TLE} generation process and whether this means that these satellites have \ac{TLE} sets with differing characteristics. The question remains, therefore, whether bias models developed using these objects can be extrapolated to other satellites and objects in the \ac{TLE} catalogue.

On the other hand, an alternative source of \enquote{ground truth} data, or more accurately, approximate \enquote{ground truth} data, might be derived from the \acp{TLE} themselves using \ac{P-OD}. By conducting long-term fits, as demonstrated in \cref{s:test_fits}, the approximate true trajectory of a given object can be extracted and then compared to the underlying \acp{TLE} to estimate their bias. This method remains restricted to passive objects due to the non-manoeuvring requirement, however due to the large number of passive objects in the \ac{TLE} catalogue, such as decommissioned satellites, rocket bodies, and other debris, it should be possible to generate a large dataset for this analysis.

\subsection{Fit Covariance Analysis}

\begin{figure}[p]
    \centering
    
    \includegraphics{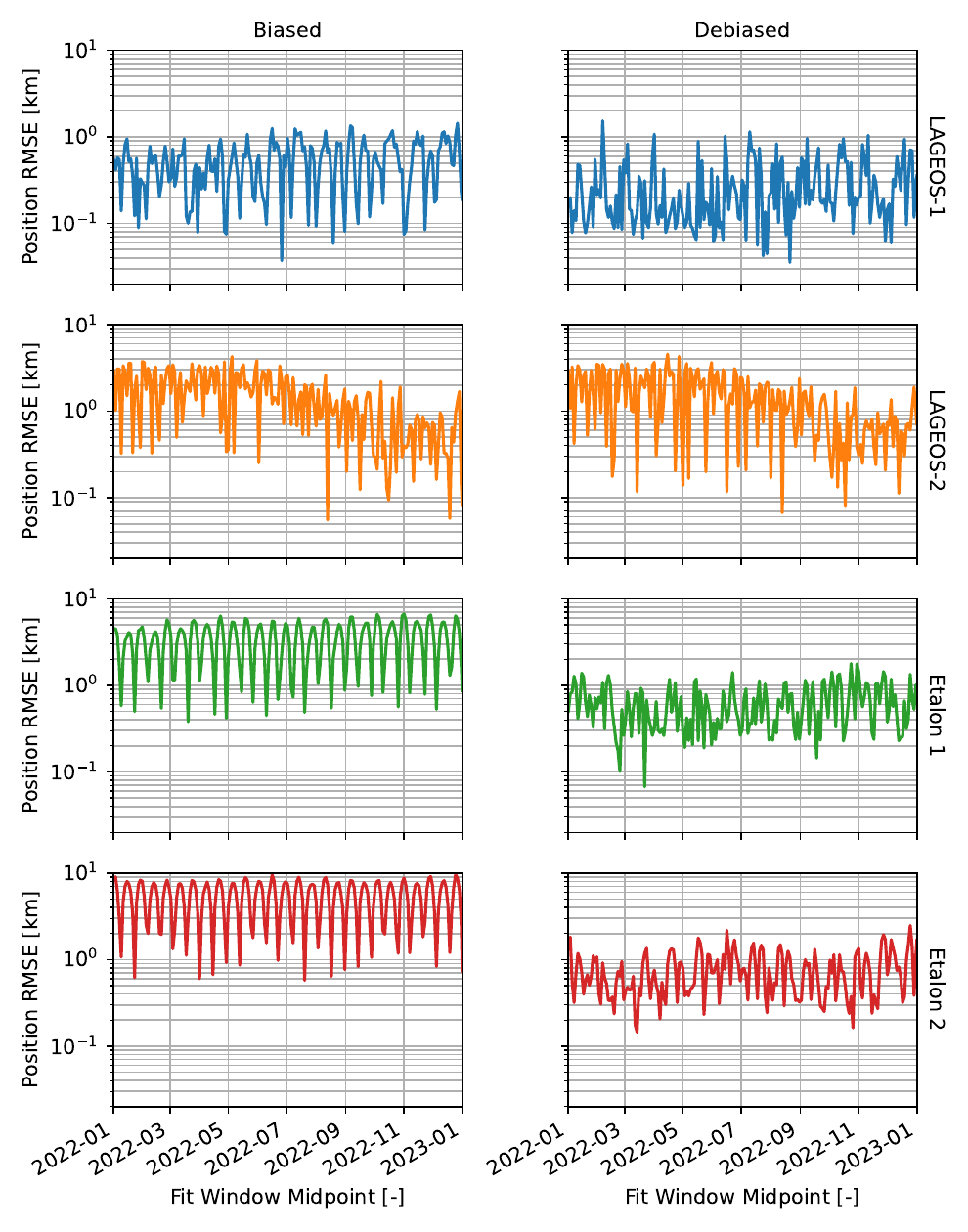}
    
    \caption{Position \acs{RMSE} during the 30~days post-fit for the \acs{LAGEOS} and Etalon satellites as a function of fit window midpoint for 10-day fits.}
    \label{f:position_rmse_portrait}
\end{figure}

\begin{figure}[p]
    \centering
    
    \includegraphics{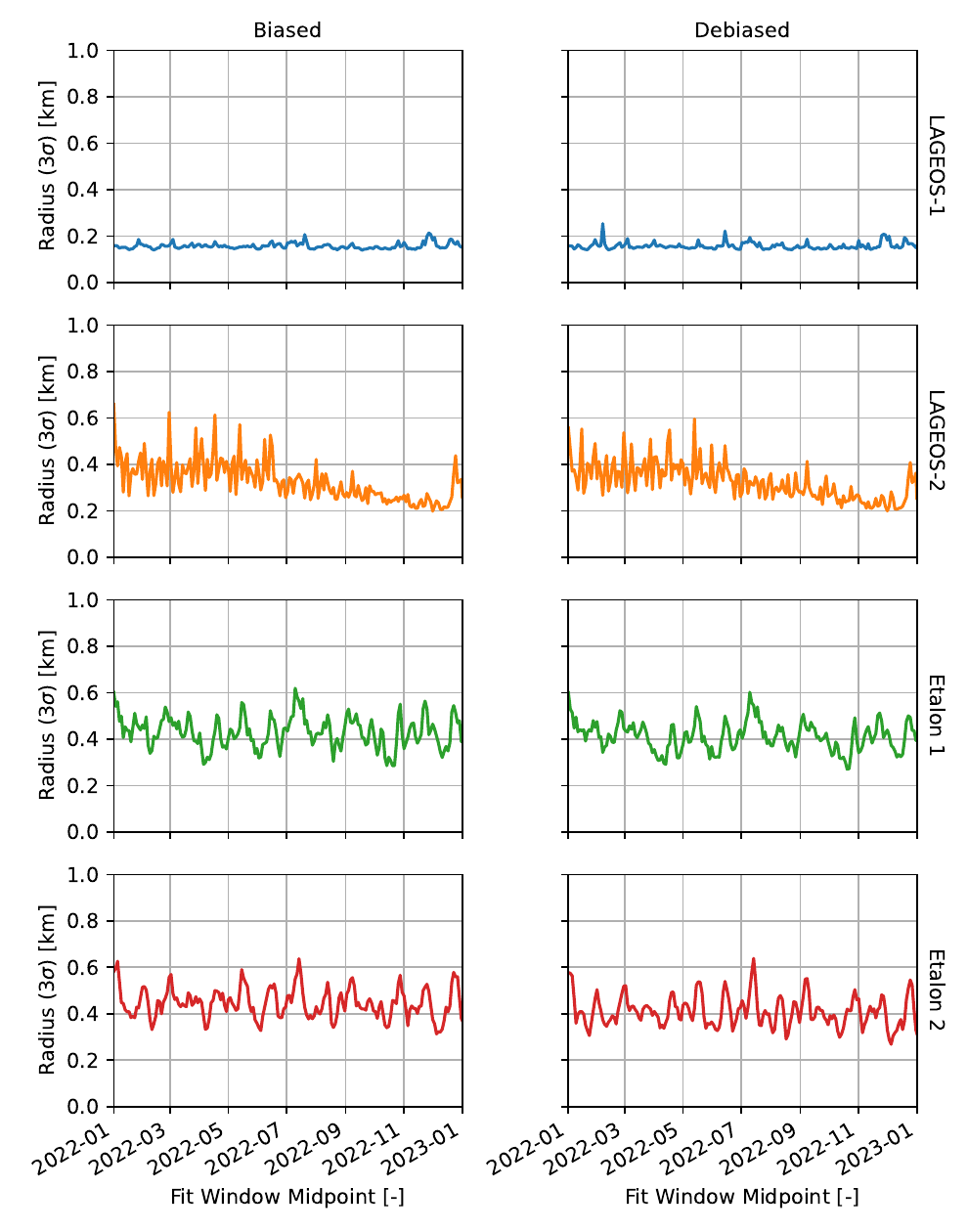}
    
    \caption{Equivalent radius of the position residuals for the \acs{LAGEOS} and Etalon satellites as a function of fit window start for 10-day fits.}
    \label{f:covariance_radius_portrait}
\end{figure}

A comparison of short-term fits conducted on the four calibration satellites, both without and with debiasing with the simplified model, are presented in \cref{f:position_rmse_portrait}. For the satellites with the best-fitting bias models, Etalon 1 and 2, debiasing had a significant impact on post-fit position \acp{RMSE}, as described in \cref{s:cal_debiased_fits}. The equivalent radii of the position residual covariances for the four calibration satellites are presented in \cref{f:covariance_radius_portrait}. In contrast to the position \acp{RMSE}, the equivalent radii throughout the year for all four objects appeared largely unaffected by the debiasing of the pseudo-observations. This suggested that the covariance alone was a poor indicator for the quality of the fit.

\begin{figure}[p]
    \centering

    \begin{subfigure}[c]{\textwidth}
        \centering
        \includegraphics{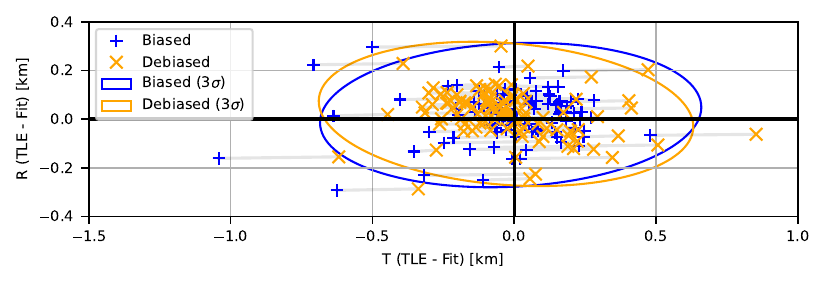}

        \caption{With respect to \acsp{TLE}.}
        \label{f:etalon1_short_scatter_tle}
    \end{subfigure}

    \begin{subfigure}[c]{\textwidth}
        \centering
        \includegraphics{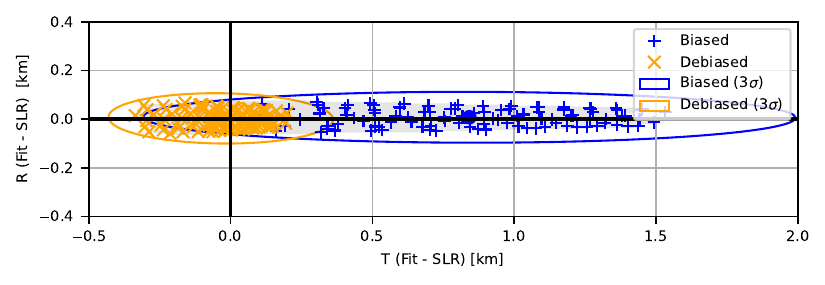}

        \caption{With respect to \acs{SLR}.}
        \label{f:etalon1_short_scatter_slr}
    \end{subfigure}    

    \caption{Example residual scatters in the radial-transverse plane for a short-term Etalon~1 fit.}
    \label{f:etalon1_short_scatter}
\end{figure}

\begin{figure}[p]
    \centering

    \begin{subfigure}[c]{\textwidth}
        \centering
        \includegraphics{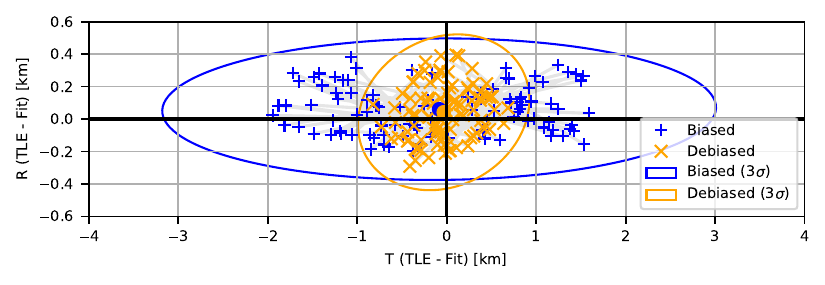}

        \caption{With respect to \acsp{TLE}.}
    \end{subfigure}

    \begin{subfigure}[c]{\textwidth}
        \centering
        \includegraphics{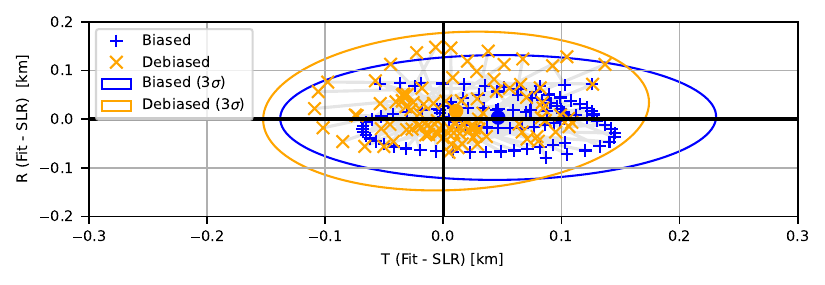}

        \caption{With respect to \acs{SLR}.}
    \end{subfigure}    

    \caption{Example residual scatters in the radial-transverse plane for a long-term Etalon~1 fit.}
    \label{f:etalon1_long_scatter}
\end{figure}

Examples of the residual scatter, the differences between the predicted and observed states at each observation time in the fit window, relative both to the \acs{TLE}-based pseudo-observations and the \acs{SLR}-based \enquote{ground truth}, for short- and long-term fits are presented in \cref{f:etalon1_short_scatter,f:etalon1_long_scatter}. The fit residuals for the short-term window, as presented in \cref{f:etalon1_short_scatter_tle}, remained largely unaffected in aggregate by the debiasing process. Although individual points shifted, in some cases by up to \SI{0.5}{\kilo\metre} in the tangential direction, the overall distribution, and hence covariance, remained approximately the same. When considering the residuals with respect to the \enquote{ground truth} in \cref{f:etalon1_short_scatter_slr}, however, a large change was visible with the both the mean of the distribution shifting towards the origin and the covariance shrinking in the tangential axis. This relationship was reversed for the case of the long-term fit window: debiasing resulted in a significant reduction in the variance along the tangential axis for the fit residuals, while the residuals with respect to the \enquote{ground truth} appeared less affected in aggregate.

The systematic variation in along-track error means that the biases in successive \acp{TLE} are highly correlated. The period of this variation is much longer than the short-term fit window sizes, therefore the pseudo-observations have similar, correlated biases. In effect, the biased \acp{TLE} are representing physically accurate and consistent orbits, however not matching the true orbits of the satellites on short time scales. The debiasing process reduces the mean error of the pseudo-observations with respect to the true state, however the noise remains unaffected. This highlights that the covariance cannot be used as an indicator for fit accuracy as it is unaffected by the biases.

\section{Conclusions}\label{s:conclusions}

Systematic along-track biases throughout the year were identified as the primary limiting factor in the quality of \ac{P-OD} fits in the \ac{MEO} regime. Mitigation of these biases was demonstrated to be key for improving the quality of higher precision ephemerides derived from \acp{TLE}. Two strategies were proposed and demonstrated: extension of the fit windows to convert short-term biases into long-term noise, and a simple sinusoidal model for estimating along-track error. Both methods were effective at reducing the typical post-fit accuracy by approaching an order of magnitude for the selected calibration satellites, and suppressing oscillations in errors for the test satellites.

It has been shown that the covariance output from the \ac{P-OD} process cannot be used to evaluate the prediction accuracy of a given fit. Many state estimation algorithms require that observations are unbiased and satisfy a given distribution, typically Gaussian, about the true states of a satellite. This requirement is not satisfied by uncorrected \acs{TLE}-based pseudo-observations during short-term fit windows, resulting in fits which appear well correlated with pseudo-observations but, in reality, are poorly correlated with the true underlying state.

The \ac{TLE} catalogue remains one of the few sources of ephemerides for a significant proportion of \acp{RSO} in near-Earth orbits. Until Type 4 \acp{TLE} are released publicly for a critical share of the catalogue, a combination of debiasing and \ac{P-OD} will be required for deriving higher-precision ephemerides for many \acp{RSO}. The development of generalised bias estimation models, covering multiple orbital regimes, will be crucial for improving \ac{P-OD} applied to the \ac{TLE} catalogue.

The analysis in this study was limited to only six passive satellites in \ac{MEO} and two in \ac{GEO}/\ac{GSO} and, therefore, may not represent general \ac{TLE} behaviour at these altitudes. There are additional calibration satellites available in this orbital regime with precision ephemerides which could be used for further investigation, namely the active satellites within various \ac{GNSS} constellations. The satellites in these constellations have similar orbits to the Etalon and tested Navstar satellites, primarily in terms of semi-major axis, eccentricity, and inclination. Nevertheless, these satellites regularly manoeuvre for station-keeping which will need to be taken into account by any analysis of their \acp{TLE}.

It is expected that \ac{TLE} behaviour will be different in \ac{LEO} due to the changes in perturbations model, notably the removal of lunisolar perturbations and \ac{SRP}, but the inclusion of atmospheric drag. A separate investigation into the biases present in this orbital regime will be required. It is expected that the same systematic biases seen with \ac{MEO} calibration satellites will not be present, instead replaced by issues with drag estimation, and sampling strategy as investigated by \cite{chen_research_2023}.

Several of the limitations of batch least squares methods can be addressed by sequential filters; therefore, these are the logical next step for improving \acs{TLE}-based \ac{P-OD}. These types of filters lend themselves naturally to solution updates as new \acp{TLE} are published by \ac{USSF}, and allow for physical parameters to be included in the estimation process continuously, such as ballistic coefficient for \ac{LEO}. The state of the art solution for sequential P-OD is to use an \ac{UKF}, initialised with an initial state and covariance from a batch least squares solution. Nevertheless, it should be noted that both batch least squares methods and sequential filters will be susceptible to biases in the pseudo-observations without the ability to detect degraded estimates through the covariance. \ac{TLE} biases must be addressed before conducting any \ac{P-OD}, particularly with methods which only consider the pseudo-observed states over a short time period.

\section*{Acknowledgments}
The authors are grateful to Alejandro Cano S\'anchez (GMV) for providing us with his perspective on interpreting batch least squares covariance results, and Creon Levit (Planet) for providing data for replication of his results.

\bibliographystyle{jasr-model5-names}
\biboptions{authoryear}
\bibliography{refs}

\end{document}